\documentclass[final,12pt,twocolumn]{elsarticle}

\usepackage{amssymb}
\usepackage{amsthm}

\usepackage{lineno}
\usepackage[hyphens]{url}
\usepackage[hidelinks]{hyperref}
\usepackage{subcaption}

\newcommand{\elem}[2]{$\mathrm{^{#1}{#2}}$}
\newcommand{\uPIC}{$\mu$-PIC}

\journal{Preprint submitted to Nucl. Instr. Meth. A}

\begin{document}

\begin{frontmatter}


\title{Development of a low-background micro pixel chamber for directional dark matter searches}



\author[inst1]{Ryota Namai\corref{cor}}
\ead{r_namai@stu.kobe-u.ac.jp}
\author[inst1]{Satoshi Higashino}
\author[inst1]{Hirohisa Ishiura}
\author[inst2]{Tomonori Ikeda}
\author[inst1]{Mizuno Ofuji}
\author[inst1]{Ayaka Nakayama}
\author[inst1]{Ryo kubota}
\author[inst7]{Kiseki D. Nakamura}
\author[inst5]{Hiroshi Ito}
\author[inst8]{Koichi Ichimura}
\author[inst3,inst4]{Ko Abe}
\author[inst6]{Kazuyoshi Kobayashi}
\author[inst2]{Atsushi Takada}
\author[inst1]{Kentaro Miuchi}
\cortext[cor]{Corresponding author}

\affiliation[inst1]{organization={Department of Physics, Kobe University},
    addressline={Rokkodai-cho 1-1}, 
    city={Kobe},
    postcode={657-8501}, 
    state={Hyogo},
    country={Japan}}

\affiliation[inst2]{organization={Department of Physics, Kyoto University},
    addressline={Kitashirakawaoiwake-cho Sakyo-ku}, 
    city={Kyoto},
    postcode={606-8502}, 
    state={Kyoto},
    country={Japan}}

\affiliation[inst3]{organization={Kamioka Observatory, Institute for Cosmic Ray Research, the University of Tokyo},
    addressline={Higashi-Mozumi 456, Kamioka}, 
    city={Hida},
    postcode={506-1205}, 
    state={Gifu},
    country={Japan}}

\affiliation[inst4]{organization={Kavli Institute for the Physics and Mathematics of the Universe (WPI), the University of Tokyo},
    addressline={Kashiwa}, 
    city={Chiba},
    postcode={277-8582}, 
    state={Chiba},
    country={Japan}}
   
\affiliation[inst5]{organization={Department of Physics and Astronomy, Faculty of Science and Technology, Tokyo University of Science},
    addressline={Yamazaki 2461}, 
    city={Noda},
    postcode={278-8510}, 
    state={Chiaba},
    country={Japan}}
    
\affiliation[inst6]{organization={Waseda Research Institute for Science and Engineering},
    addressline={Okuba 3-4-1}, 
    city={Shinjuku},
    postcode={169-8555}, 
    state={Tokyo},
    country={Japan}}            
     
\affiliation[inst7]{organization={Department of Physics, Graduate School of Science and Faculty of Science, Tohoku University},
    addressline={6-3, Aramaki Aza-Aoba, Aoba-ku}, 
    city={Sendai},
    postcode={980-8578}, 
    state={Miyagi},
    country={Japan}}

\affiliation[inst8]{organization={Research Center for Neutrino Science, Tohoku University},
    addressline={6-3, Aramaki Aza-Aoba, Aoba-ku}, 
    city={Sendai},
    postcode={980-8578}, 
    state={Miyagi},
    country={Japan}}

\begin{abstract}
Direct detection of weakly interacting massive particles (WIMPs) can provide strong evidence of their existence and the directional method would have an advantage over other methods to detect the clear signal of WIMPs. Time projection chambers with micro-patterned gaseous detectors (MPGDs) are one of the common devices used in directional WIMP searches. A micro pixel chamber (\uPIC), one of the various types of MPGDs, with specially selected low background materials (LBG\uPIC) was developed and its performance was studied. The radon emission of the LBG\uPIC\ was less than 1/60 of that of the \uPIC\ currently in use. Although a non-negligible gain non-homogeneity was seen for the LBG\uPIC, it can be used for the directional WIMP search with the correction of the non-homogeneity.
\end{abstract}



\end{frontmatter}

\section{Introduction}
\label{sec:introduction}
A number of observational results at various cosmological scales which are not explained with known physics can be understood by introducing an undiscovered new particle, the dark matter. 
Weakly Interacting Massive Particles (WIMPs) are ones of the strong dark matter candidates with a long search history. Although a number of direct WIMP search experiments have been conducted, they have not reached to its discovery. Directional methods sensitive to the anisotropy of the recoil nuclei are said to provide clear signatures of WIMPs. This anisotropy is caused by the move of the solar system in our galaxy which makes a large anisotropy in the incoming directions of the WIMPs towards the detectors set on the Earth. Time projection chambers (TPCs) with micro-patterned gaseous detectors (MPGDs) can detect three dimensional tracks of the recoil nuclei and are widely used for the directional WIMP searches~\cite{AHLEN2011124, BATTAT201765,Shimada:2023vky}. 

\begin{figure*}[htbp]
    \centering
    \includegraphics[width=\textwidth]{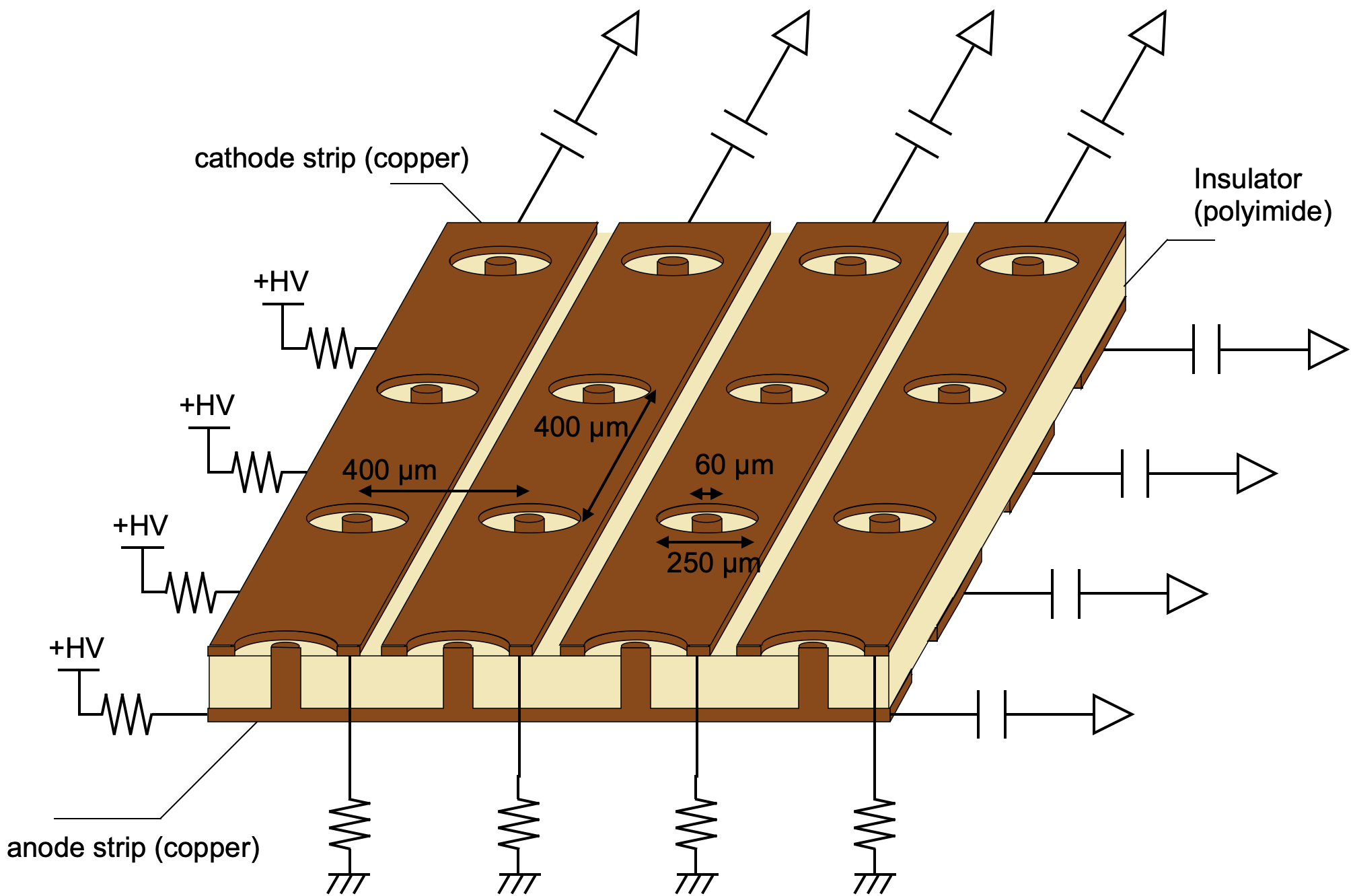}
    \caption{Schematic of \uPIC}
    \label{fig:la_img}
\end{figure*}

Micro pixel chambers (\uPIC s), ones of the MPGDs, have been used as the readout device for the gaseous TPC in the NEWAGE experiment~\cite{Shimada:2023vky}. The \uPIC\ is a two-dimensional strip-readout detector with a pitch of $\mathrm{400\ \mu m}$ having a typical detection area of $\mathrm{30.72 \times 30.72~cm^2}$. The \uPIC\ fabricated with the printed circuit board technology. The WIMP-search sensitivity of NEWAGE had been limited by the alpha rays originating from the radioactive isotopes \elem{238}{U} and \elem{232}{Th} in glass fibers used to reinforce the \uPIC\ board. This \uPIC\ of the original structure used in our work published in 2015~\cite{Nakamura:2015iza} is refereed as ``standard \uPIC'' in the following discussions. A new \uPIC\ without glass fibers in the surface material (``LA\uPIC'') to realize less alpha-ray emission was developed~\cite{Hashimoto:2020xly}. The LA\uPIC\ improved the sensitivity of NEWAGE, indicating that the next serious background source would be radioactive radon isotopes (\elem{220}{Rn} and \elem{222}{Rn})~\cite{Ikeda:2021ckk}. \elem{220}{Rn} and \elem{222}{Rn} are the progenies of long-lived radioactive isotopes \elem{228}{U} and \elem{232}{Th} which are contaminated in most of the materials on the Earth. Radon is a noble gas and can emanate from the detector components. Their decays in the detection volume are common backgrounds for rare event search experiments. The LA\uPIC\ was found to be one of the main sources of radons which originate from the glass fibers used as inner support material and the solder resist used for insulation. We, therefore, developed a new \uPIC\ with low radioactive materials (LBG{\uPIC}) for less radon emanation and less other radiation emissions. 

In this paper, a development of the LBG\uPIC\ and its performance tests are described. The material selection and fabrication are described in Section~\ref{sec:production}. Section~\ref{sec:performance} describes the detector performance tests then the results are discussed in Section~\ref{sec:discussion}. The work is concluded in Section~\ref{sec:conclusions}.
\section{Design and Production}
\label{sec:production}
This section describes the requirements, material selections, and productions of the LBG\uPIC.

\subsection{Requirements} 
\label{sec:requirements}
The primary requirement was set to the radioactive contamination levels. Isotopes of \elem{238}{U} and \elem{232}{Th} are contaminated in most of the materials used for the detector production and rare gas radon isotopes (\elem{222}{Rn} and \elem{220}{Rn}) in their decay chains emanate out of the detector components. Their decays are associated with alpha rays, which are one of the main background sources of our experiment. We required that the contamination of these isotopes in the LBG\uPIC\ is less than one-tenth of that of the LA\uPIC~\cite{Hashimoto:2020xly}. This provides a sensitivity of WIMP search one order of magnitude better than the latest NEWAGE result~\cite{Shimada:2023vky}, assuming ten times larger exposure than the previous work. Since the rate of surface alpha-ray emission had already been reduced in the production of the LA\uPIC~\cite{Hashimoto:2020xly}, it should be retained to the level equal to or less than that of the LA\uPIC\ for this development. 

The detector performance should be equivalent to the LA\uPIC. The LA\uPIC\ has 768~$\times$~768 pixels in an area of 30.72~$\times$~30.72~cm$^{2}$, with a pitch of 400~$\mu$m. The readout pitch of LBG\uPIC\ needed to be 400~$\mu$m to have an equivalent position resolution to LA\uPIC. Since the LA\uPIC~was operated in $\mathrm{CF_4}$ gas at $76~\mathrm{Torr}$ in our previous WIMP search~\cite{Shimada:2023vky} with a gas gain of 1000 or more, we required the LBG\uPIC\ to show a similar gas gain in the same gas condition. Similarly, the gain uniformity of less than 20\% RMS was required for the overall detection area.

\subsection{Material}
\label{sec:material_screenings}
\subsubsection{Material Production}
The main sources of the radon emanation in LA\uPIC\ are the polyimide with Glass Cloth (PI w/~GC) used for the core material and the solder resist on the surface (see Fig.~\ref{fig:schematic_diagrams}). While the amount of solder resist can be easily reduced by redesigning, alternative materials for the core material must be found. It was known that Shin-Etsu Chemical. Co, Ltd. produces high-purity quartz cloth for high-frequency band applications~\cite{ShinnetsuSQX}. A specialized insulating material was prepared through a discussion with the production company of the \uPIC, DNP (Dai Nippon Printing Co., Ltd.), taking account of the heat process during the \uPIC\ production. In this subsection, we describe the results of the radon emanation and surface alpha-ray measurements of this new material, Quartz with Resin (Quartz w/~Resin).

\subsubsection{Gamma-ray Measurements}
\label{sec:internal_alpha rays}
Radioisotope contamination in the candidate materials was investigated with a high purity Germanium (HPGe) detector at Kamioka Observatory~\citep{XMASS:2018saz}. Contamination amounts  of \elem{238}{U} and \elem{232}{Th} radioactive chains were of particular interest as sources of the radon isotopes and $\gamma$-ray rates in these chains were measured. Measured results were interpreted as the equivalent contamination amounts of \elem{238}{U} (``upper'' stream in the \elem{238}{U}), \elem{226}{Ra} ( ``middle'' stream in the \elem{238}{U}), and \elem{232}{Th}. The upper and middle streams of the \elem{238}{U} were evaluated separately because it is known that the radioactive equilibrium is broken in many cases. The results are shown in Table~\ref{tab:hpge_result}. The U/Th contamination levels of the new material candidate, Quartz w/~Resin were found to be less than $10^{-2}$ of those of the previous one, PI w/~GC. This result indicates that Quartz w/~Resin satisfies our requirement from the viewpoint of the radon emanation sources.

\begin{table*}[htbp]
    \centering
    \begin{tabular}{c|c|c|c}
        Sample & \elem{238}{U} upper & \elem{238}{U} middle & \elem{232}{Th} \\ 
        \hline
        \hline
        PI(w/~GC) & $(7.8 \pm 0.1) \times 10^{-1}$ & $(7.6 \pm 0.1) \times 10^{-1}$ & $3.42 \pm 0.03$ \\  
        Solder resist & $(3.9 \pm 0.1) \times 10^{-1}$ & $  < 2.3 \times 10^{-3}$ & $(4.2 \pm 0.1) \times 10^{-2}$ \\
        \hline 
        Quartz w/~Resin & $(5.6 \pm 5.2) \times 10^{-3}$ & $(5.1 \pm 1.0) \times 10^{-3}$ & $(1.2 \pm 0.4) \times 10^{-2}$ \\
        \hline
    \end{tabular}
    \caption{Results of radioactive contamination measurement with an HPGe detector. The unit is ppm($\mathrm{10^{-6}~g/g}$). The new core material candidate, Quartz w/~Resin, shows a reduction in U/Th contamination by approximately $10^{-2}$ compared to the current core material, PI w/~GC.} 
    \label{tab:hpge_result}
\end{table*}

\subsubsection{Surface Alpha-ray Measurements}
\label{sec:surface_alpha rays}
Alpha-ray emission rates from the material surfaces were measured with a surface alpha-ray counter; Ultra-Lo~1800~\citep{UltraLo1,UltraLo2,Nakib:2013ffd,McNally:2014eka} in Kamioka Observatory. Details and techniques for the related low background measurements, established by the XMASS experiment, are described in Ref.~\citep{Kobayashi:2018wlh}. Two of the main surface materials, the copper sheet and the Quartz w/~Resin, were measured. The expected surface alpha-ray rate of the LBG\uPIC\ was then calculated from these measurements. The screening results are shown in Table~\ref{tab:surface_alpha_result}. The alpha-ray emission rate of the LA\uPIC\ is shown for reference~\cite{Hashimoto:2020xly}. Expected surface alpha-ray rate of the LBG\uPIC\ was estimated to be the equivalent level of the rate of the LA\uPIC\ within errors. Based on the above results, we decided to use Quartz w/~Resin for the LBG\uPIC\ production.

\begin{table*}[htbp]
    \centering
    \begin{tabular}{c|c}
        Sample & emmissivity~$\mathrm{[\alpha/cm^{2}/hr]}$ \\
        \hline
        \hline
        LA\uPIC &  $(2.35 \pm 0.48) \times 10^{-4}$ \\
        \hline
        Copper sheet & $(1.14 \pm 0.36) \times 10^{-4}$ \\
        Quartz w/~Resin & $(2.29 \pm 0.36) \times 10^{-4}$ \\
        LBG\uPIC(expected) & $(1.63 \pm 0.51) \times 10^{-4}$ \\
        \hline
    \end{tabular}
    \caption{Results of surface alpha-rays measurement with an Ultra-Lo~1800. LBG\uPIC(expected) is the expected rate calculated with the rates of copper sheet and Quartz w/~Resin. The surface alpha rate of the LBG\uPIC\ is the equivalent level of that of the LA\uPIC\ within errors.}
    \label{tab:surface_alpha_result}
\end{table*}

\subsection{Design and production} 
\label{sec:design_and_production}
\begin{figure*}[htbp]
    \begin{subfigure}{0.5\textwidth}
        \includegraphics[width=1\linewidth]{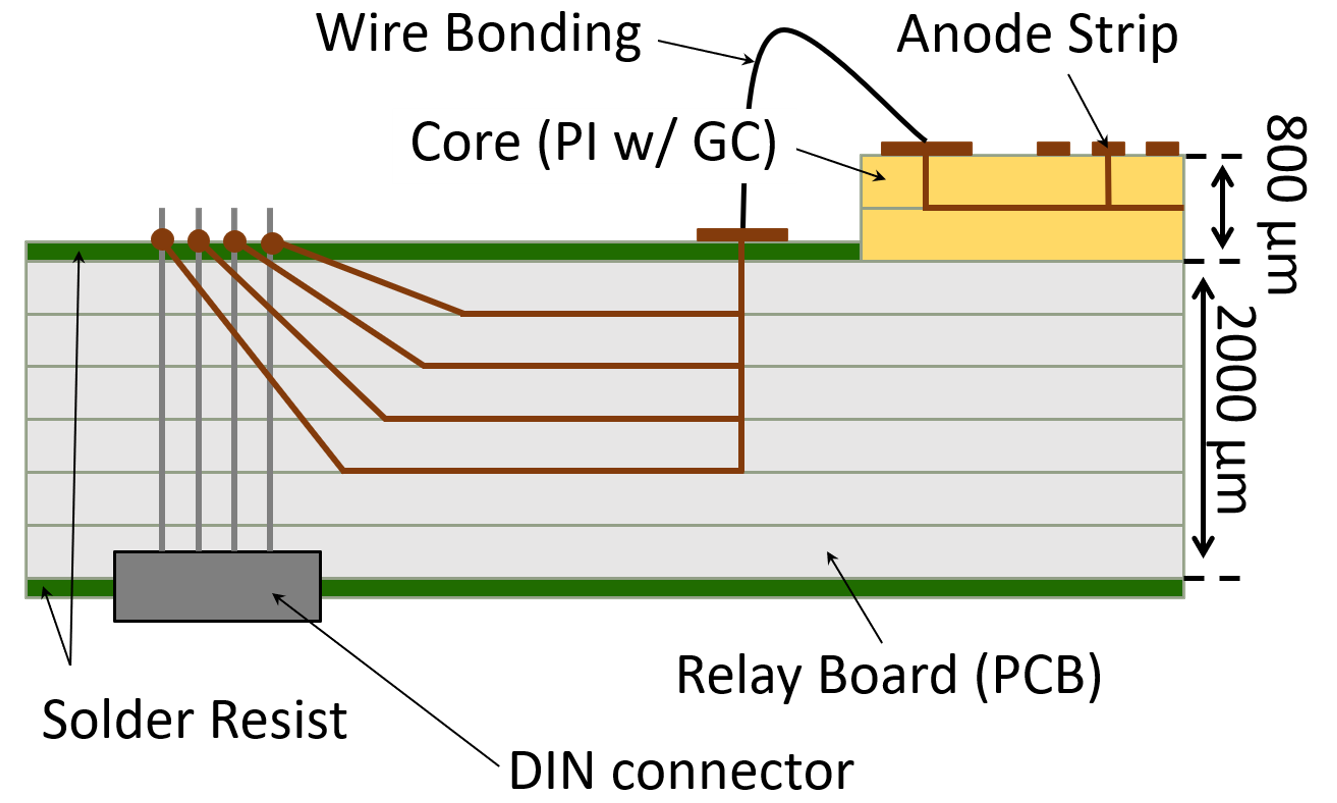}
        \caption{LA\uPIC\ schematic diagram}
        \label{fig:LAuPIC_diagram}
    \end{subfigure}
    \begin{subfigure}{0.5\textwidth}
        \includegraphics[width=1\linewidth]{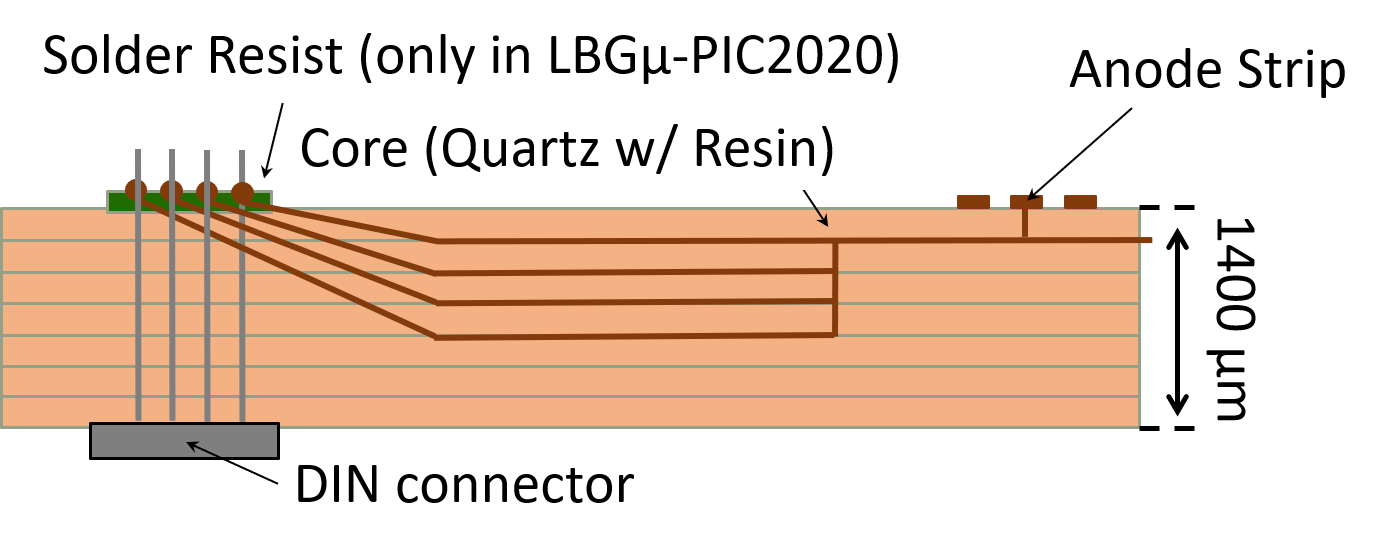}
        \caption{LBG\uPIC\ schematic diagram}
        \label{fig:LBGuPIC_diagram}
    \end{subfigure}
    \caption{Schematic cross-section drawings of the LA\uPIC\ and the LBG\uPIC. The LA\uPIC\ was made of two components; the detector part (shown with yellow layers) and the relay board (shown with light-gray layers). The LBG\uPIC\ is designed in one board of six layers (orange).}
    \label{fig:schematic_diagrams}
\end{figure*}

The LBG\uPIC\ was designed identically to the LA\uPIC\ as a two dimensional detector; a detection area of $30.72~\times~30.72~\mathrm{cm^2}$ with $768~\times~768$ pixels of a $400~\mu$m-pitch. Schematic cross-section drawings of the LA\uPIC\ and the LBG\uPIC\ are shown in Fig.~\ref{fig:schematic_diagrams}. The LA\uPIC\ was made of two components; the detector part (shown with yellow layers) and the relay board (shown with light gray layers). In order to reduce the material mass, the detector part and relay board were integrated in the LBG\uPIC\ production. This integration also removes the wire-bonding assembly process which has a risk of radioactive source contamination.

\begin{figure*}[htbp]
    \begin{tabular}{ccc}
        \begin{minipage}{.33\textwidth}
                \centering
                \includegraphics[height=4cm]{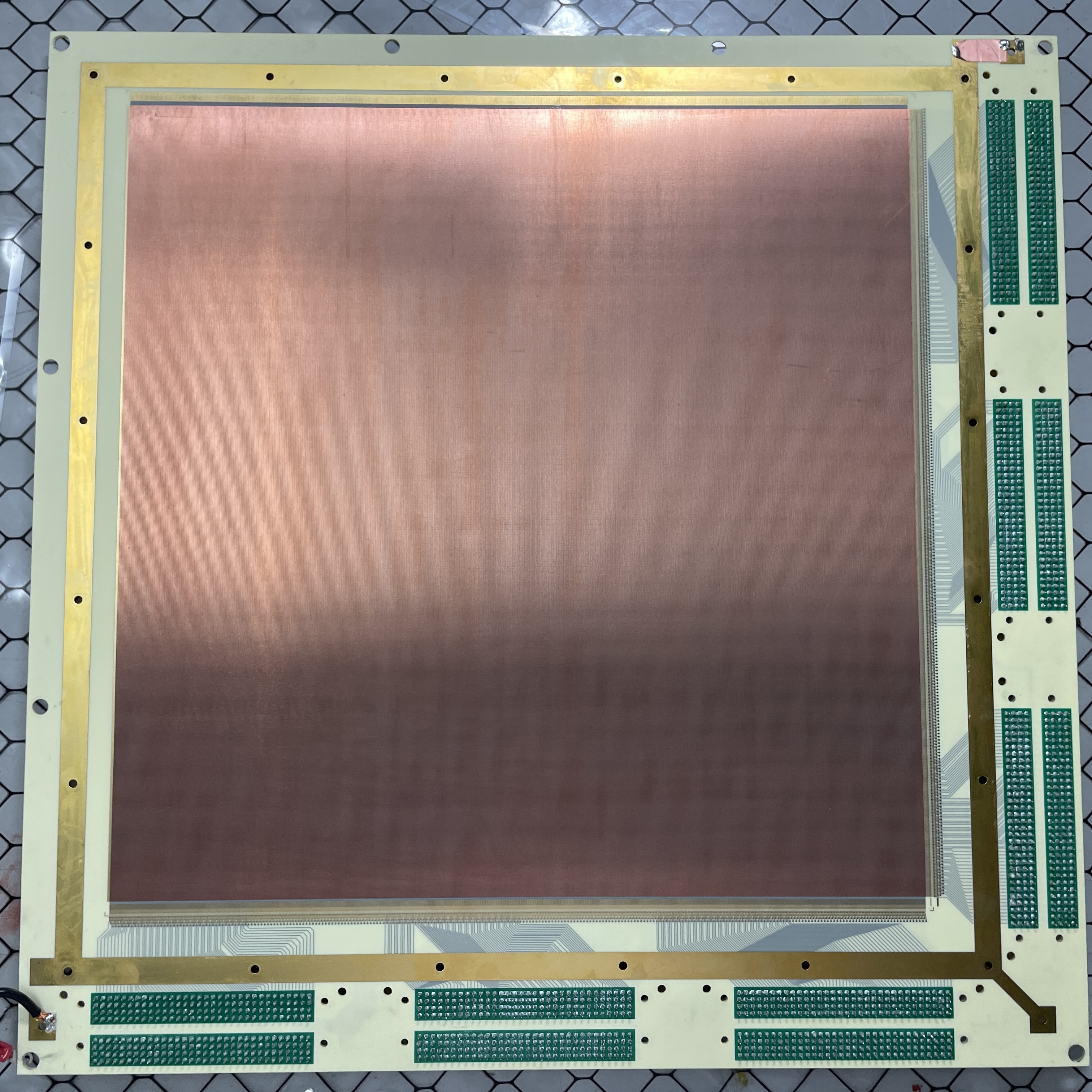}
                \subcaption{LBG\uPIC2020}
                \label{fig:lbg1_outer}    
        \end{minipage}
        \begin{minipage}{.33\textwidth}
            \centering
            \includegraphics[height=4cm]{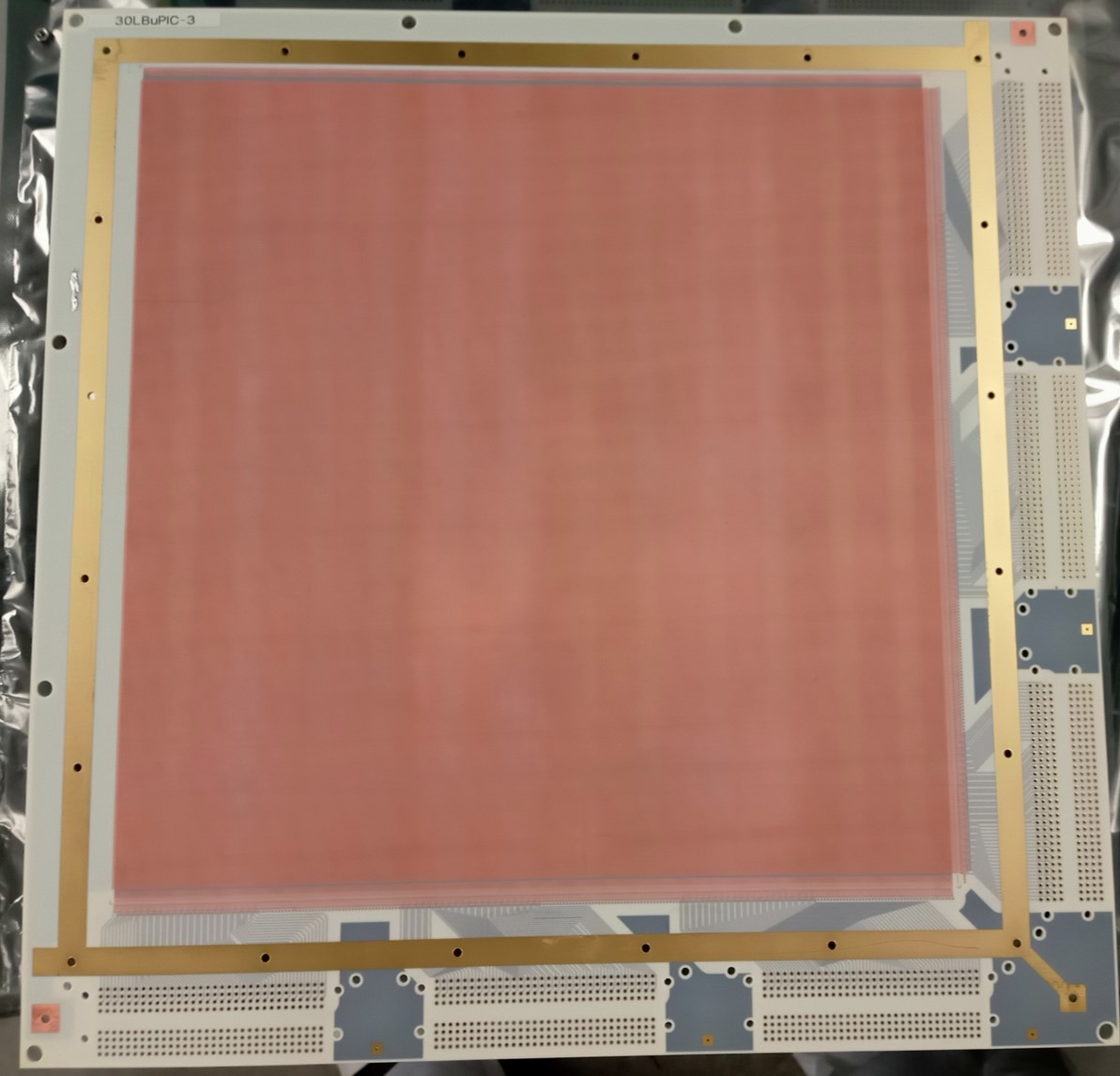}
            \subcaption{LBG\uPIC2023-1}
            \label{fig:lbg3_outer}
        \end{minipage}
        \begin{minipage}{.33\textwidth}
            \centering
            \includegraphics[height=4cm]{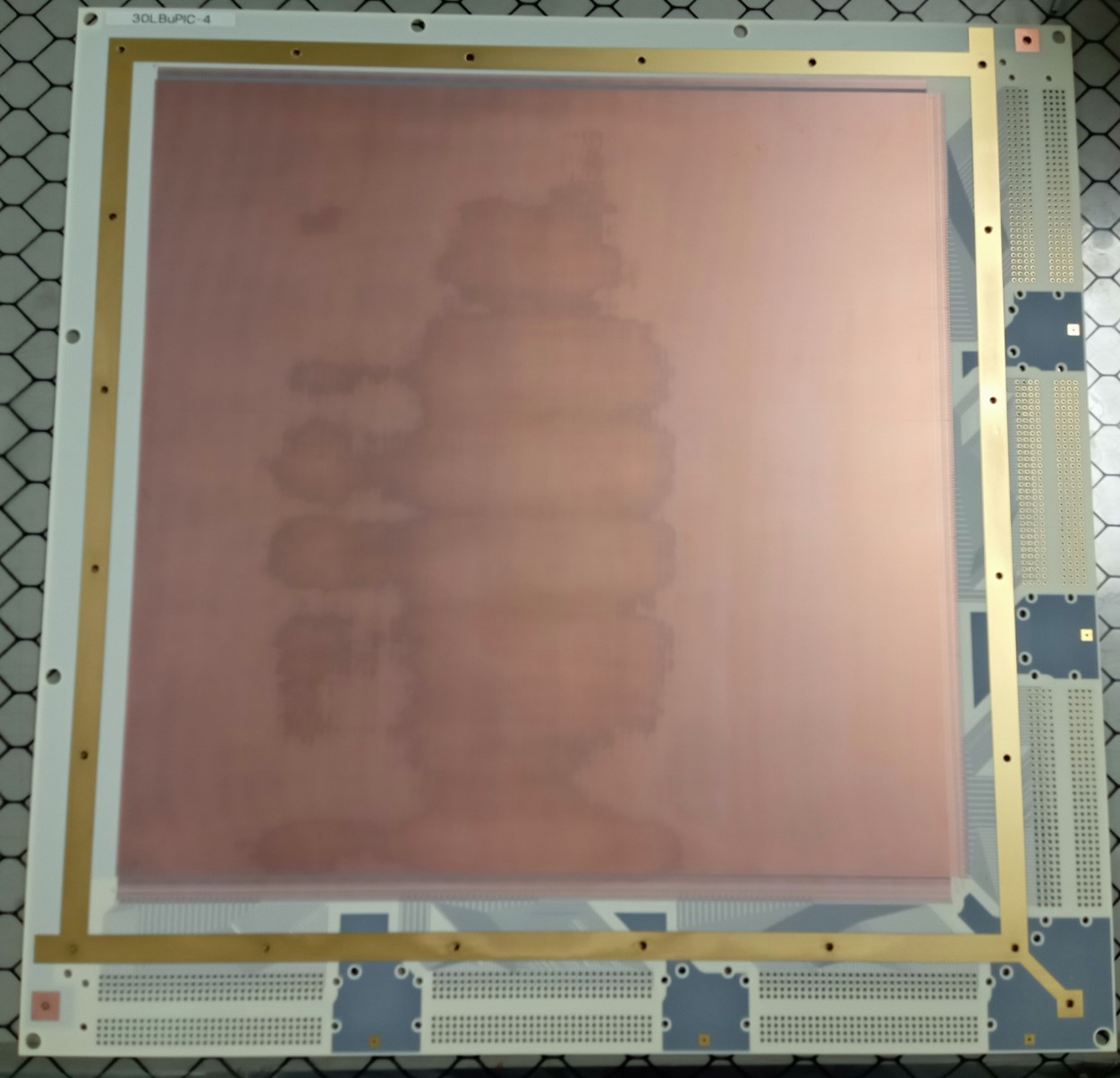}
            \subcaption{LBG\uPIC2023-2}
            \label{fig:lbg4_outer}
        \end{minipage}
    \end{tabular}
    \caption{External view of each \uPIC.}
    \label{fig:lbg_outer}
\end{figure*}

The LBG\uPIC s were manufactured by DNP. One LBG\uPIC\ (LBG\uPIC2020) was produced in 2020 (Fig.~\ref{fig:lbg1_outer}) and two LBG\uPIC s (LBG\uPIC2023-1 and LBG\uPIC2023-2) were produced in 2023 after some refining in the procedure (Fig.~\ref{fig:lbg3_outer} and~\ref{fig:lbg4_outer}). The main modification from LBG\uPIC2020 to LBG\uPIC2023s is the improvement of the plating conditions which affect the surface conditions of the electrodes. Solder resist, used on the whole connector area in LA\uPIC\ and on the limited area in LBG\uPIC2020, was not used in LBG\uPIC2023s.

\section{Performance} 
\label{sec:performance}
Since the gas gain at each pixel can be predicted by its structure, the electrodes of the LBG\uPIC\ were first visually inspected as described in Section~\ref{sec:visual_inspections}. The actual gas gains and their uniformities were then measured as described in Section~\ref{sec:gas_gain_measurement}. Finally, the amount of radon emanation and the surface alpha-ray emission from LBG\uPIC\ were measured as discussed in Section~\ref{sec:background_measurement}.

\subsection{Visual Inspections}
\label{sec:visual_inspections}
Two ways of visual inspections,
detailed and all-area 
visual inspections were carried out to
quantify the formations of the electrodes.

\subsubsection{Detailed Visual Inspection}
\label{sec:partial_visual_inspections}
\begin{figure*}[htbp]
    \begin{tabular}{cc}
        \begin{minipage}{.48\textwidth}
                \centering
                \includegraphics[height=8cm]{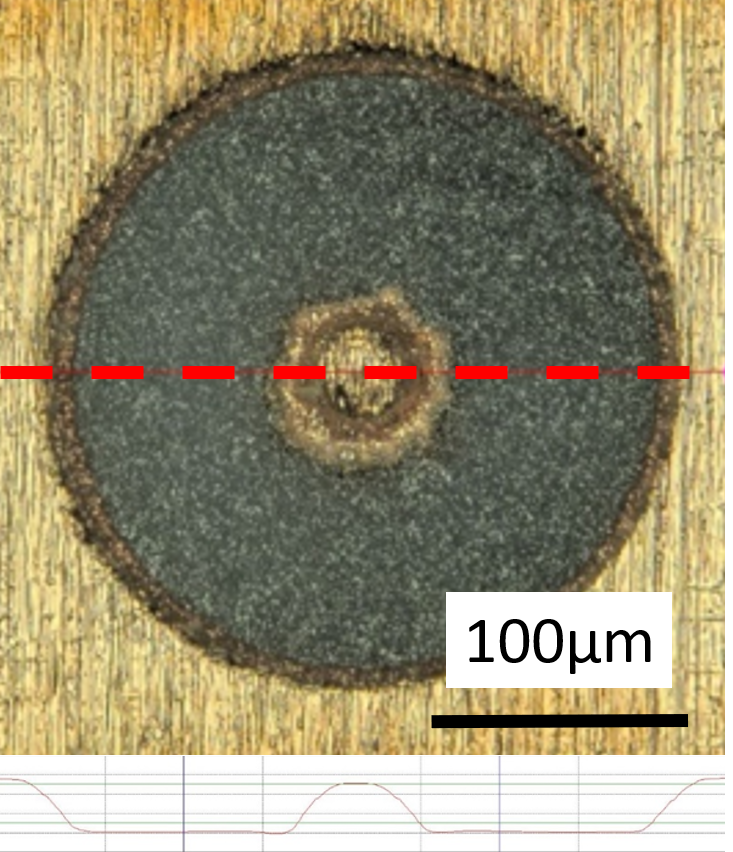}
                \subcaption{Pixel formation is normal part}
                \label{fig:pixel_good}
        \end{minipage}
        \hspace{.04\textwidth}
        \begin{minipage}{.48\textwidth}
            \centering
            \includegraphics[height=8cm]{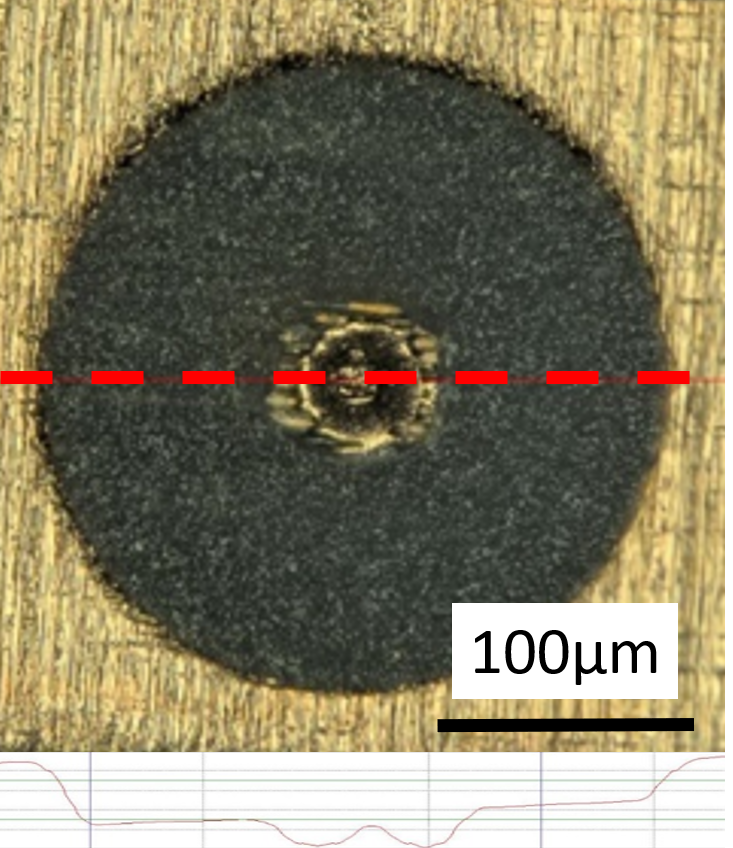}
            \subcaption{Areas where pixel formation is not normal}
            \label{fig:pixel_bad}
        \end{minipage}
    \end{tabular}
    \caption{Photographs of LBG\uPIC2020 taken with the VHX-2000 microscope at $800\times$ magnification. The lower panels show cross-sections of the electrodes along the dashed red lines in the images above. (a) Photograph of a normally formed pixel.  (b) Photograph of a poorly formed pixel. The top of the anode electrode in (a) is above the surrounding insulator while that in the right panel is the same or below the surrounding insulator.}
    \label{fig:pixel_example}
\end{figure*}

Anode and cathode electrodes in a limited area were visually inspected in detail with an optical microscope (Keyence, VHX-2000). This microscope can take a three-dimensional image of the object, although the working area is limited to 10 $\times$ 10 $\mathrm{cm}^2$.\ Typical images of the electrodes taken with this microscope are shown in Fig.~\ref{fig:pixel_example}. The upper panels show two-dimensional top-view images, and the lower panels show the cross sections of the images above along the dashed red lines. Fig.~\ref{fig:pixel_good} shows an image of a well-formed electrode, while that of a poorly-formed electrode is shown in Fig.~\ref{fig:pixel_bad}. The central part of each cross section is the positions of the anode electrode where the large difference between these two images is seen; the top of the anode electrode in the left panel is above the surrounding insulators while that in the right panel is the same or below the surrounding insulators. Anode electrodes are formed by plating process and it is known that there occurs some insufficient growing due to the technical difficulties of the process. While this method provides us with three-dimensional structural information of the electrodes sufficient to predict the gas gains by simulations, the inspection area is limited as aforementioned. In order to address these issues, we developed a new method to quantify the formation of all pixel electrodes, which is described in the next section.

\subsubsection{Whole-area Visual Inspection}
\label{sec:overall_visual_inspection}
\begin{figure*}[htbp]
    \begin{tabular}{ccc}
        \begin{minipage}{.48\textwidth}
            \centering
            \includegraphics[height=5cm]{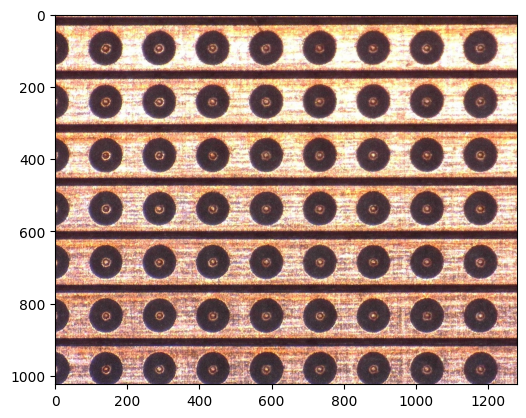}
            \subcaption{Original image}
            \label{fig:detect_org}
        \end{minipage}
        \begin{minipage}{.48\textwidth}
            \centering
            \includegraphics[height=5cm]{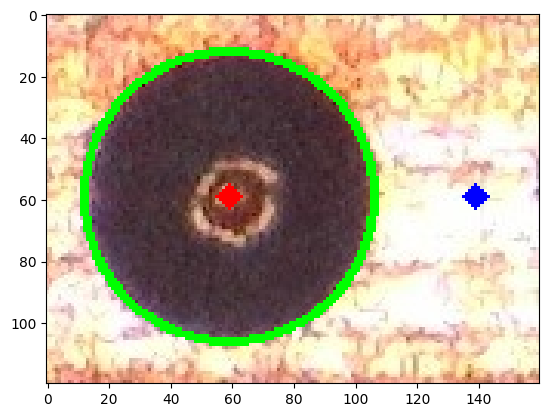}
            \subcaption{Enlarged image}
            \label{fig:detect_det}
        \end{minipage}
    \end{tabular}
    \caption{Analysis method on the pictures of the electrode. (a) Original image. (b) Enlarged image for one pixel for $R_{\rm{C}}$, $Br_{\rm{A}}$ and $Br_{\rm{C}}$ detection. The detected circle, reference point of the anode electrode, and that of the cathode electrode are shown by a green circle, a red point and a blue point, respectively.}
    \label{fig:detect_img}
\end{figure*}

The inspection for the whole area was carried out with a visual inspection system consisting of a USB microscope and a computer numerically controlled (CNC) X-Y stage. The system was driven by Python scripts allowing to position the USB microscope for an area of $\mathrm{40\times 36~cm^2}$, and to take pictures with a variable magnification of between 10 and 140. Pictures were taken at a rate of about $\mathrm{1000~cm^2/day}$ with a magnification of 116. An example of an image, typically containing $8\times7$ electrodes is shown in Fig.~\ref{fig:detect_org}. Two parameters, cathode radius $R_{\rm C}$ and brightness $Br$ are closely related to the gas gain. If the anode height is lower than normal, the gain will be lower, and the anode will appear darker in the picture. On the other hand, $R_{\rm C}$ is small, the electric field generated in the strip is larger, so the gain is larger. Each parameter was measured in the following way.
\begin{enumerate}
    \item Detect the cathode circle using Open-CV's HoughCircle function on the pixel of interest (shown by the green circle in Fig.~\ref{fig:detect_det}). The radius of this circle is defined as $R_{\rm C}$.
    \item Define the center of the cathode circle as the reference point of the anode electrode (shown by the red point in Fig.~\ref{fig:detect_det}). Also, the position 100 pixels apart in the x-direction from the reference point of the anode electrode is defined as the reference point of the cathode electrode (shown by the blue point in Fig.~\ref{fig:detect_det}).
    \item Convert the image to 8-bit grayscale and calculate the brightness of the anode and cathode electrodes ($Br_{\rm{A}}$ and $Br_{\rm{C}}$) by averaging the values within 3~pixels of radius, which is sufficiently smaller than the anode radius, from the reference points.
\end{enumerate}
After extracting values of $Br_{\rm{A}}$ and $Br_{\rm{C}}$ for each pixel, $Br$ was calculated by the following equation
\begin{equation}
    Br = \frac{Br_{\rm{A}}-Br_{\rm{C}}}{Br_{\rm{C}}}.
    \label{eq:brightness}    
\end{equation}

\begin{figure*}[htbp]
    \begin{tabular}{ccc}
        \begin{minipage}{.32\textwidth}
            \centering
            \includegraphics[height=3.8cm]{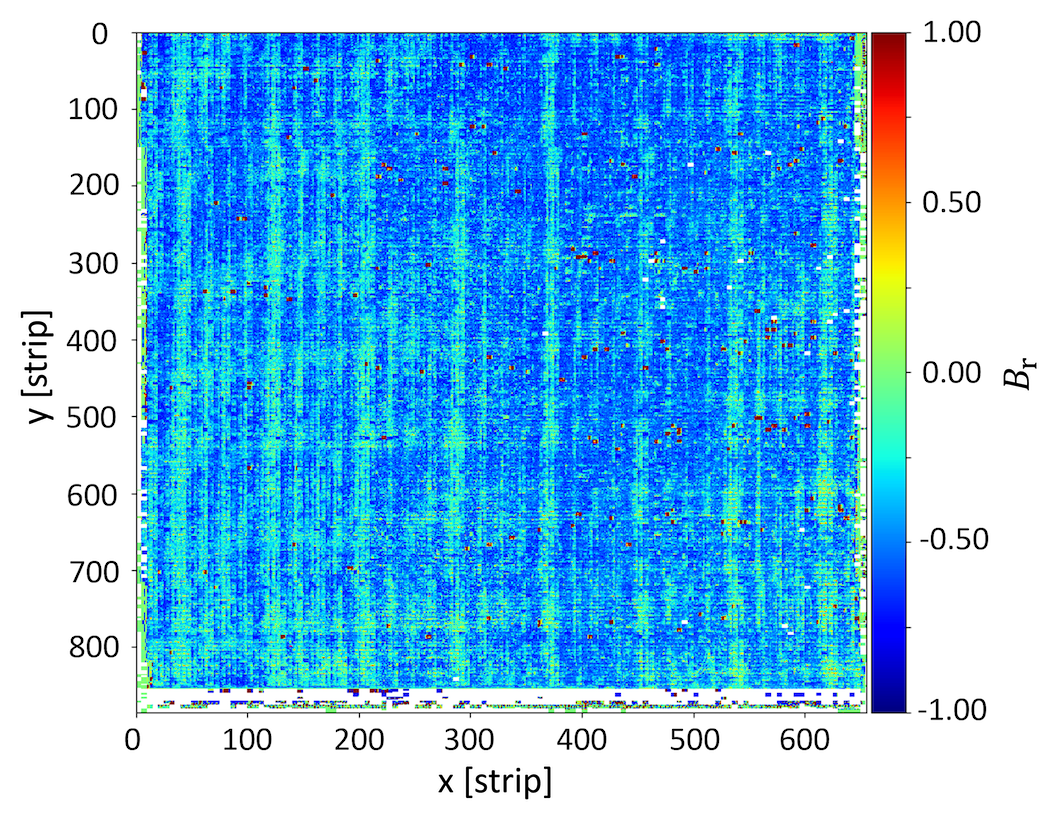}
            \subcaption{LBG\uPIC2020}
            \label{fig:lbg1_norm2d}
        \end{minipage}
        \hspace{.01\textwidth}
        \begin{minipage}{.32\textwidth}
            \centering
            \includegraphics[height=3.8cm]{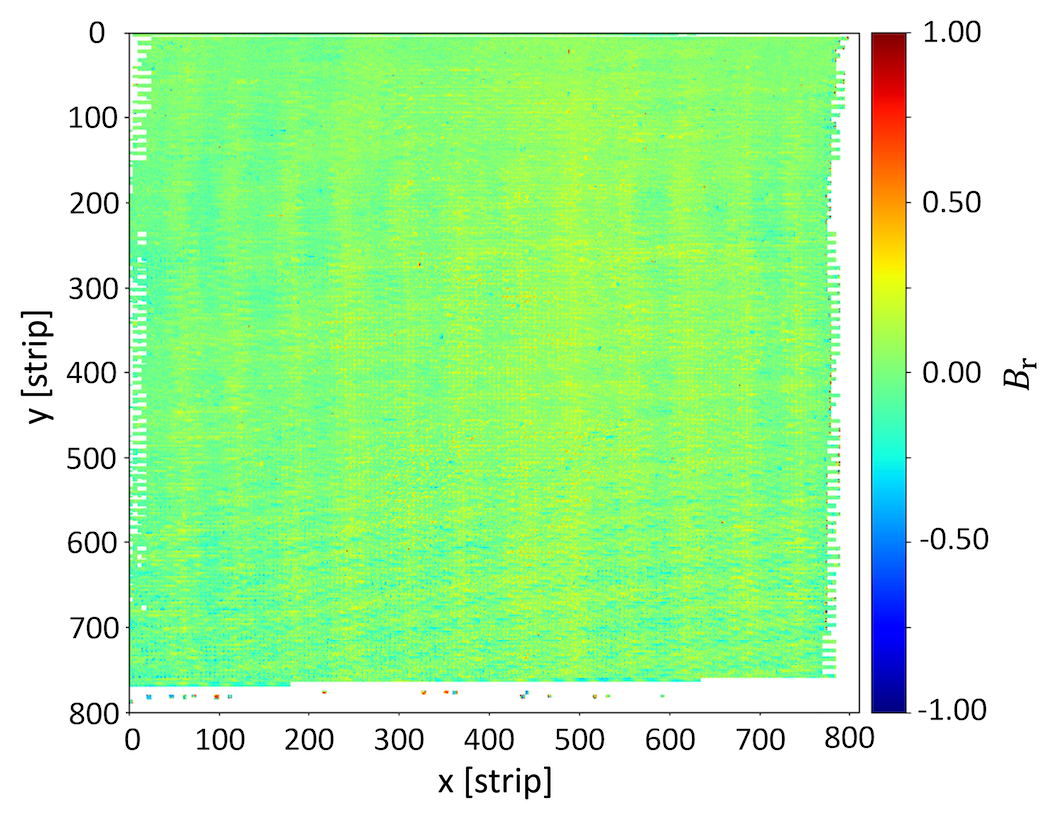}
            \subcaption{LBG\uPIC2023-1}
            \label{fig:lbg3_norm2d}
        \end{minipage}
        \hspace{.01\textwidth}
        \begin{minipage}{.32\textwidth}
            \centering
            \includegraphics[height=3.8cm]{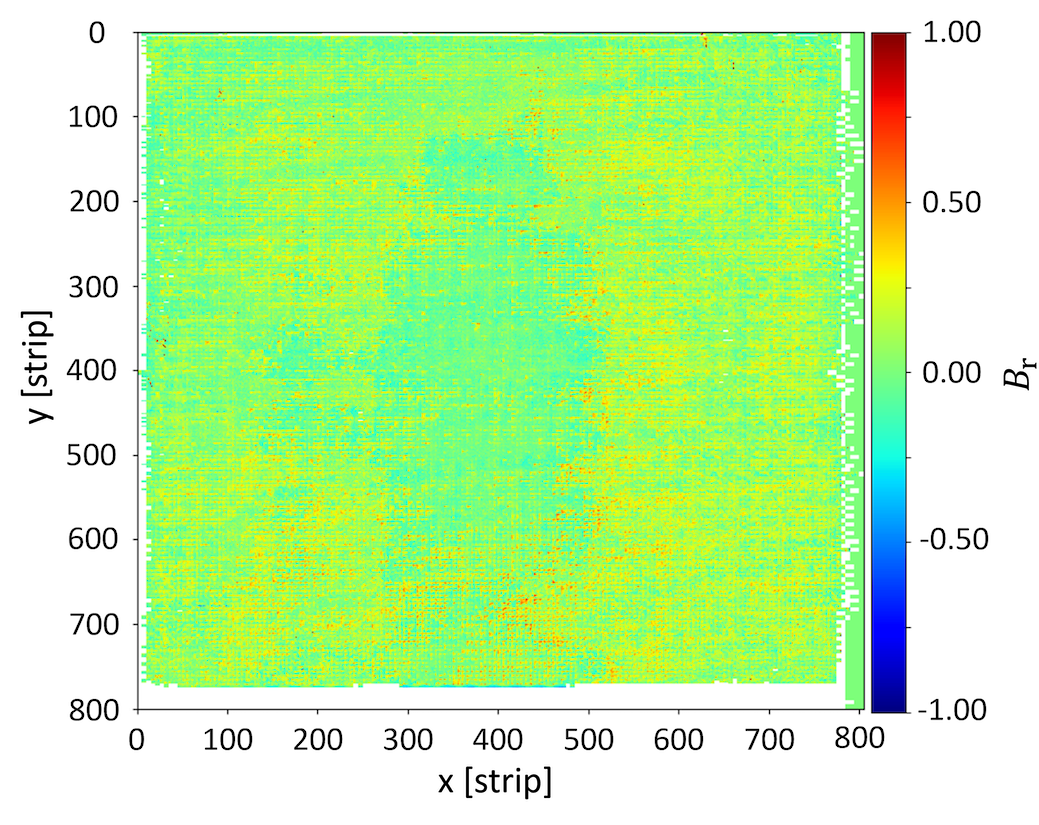}
            \subcaption{LBG\uPIC2023-2}
            \label{fig:lbg4_norm2d}
        \end{minipage}
    \end{tabular}
    \caption{Two-dimensional distributions of $Br$.}
    \label{fig:norm2d}
\end{figure*}

\begin{figure}[htbp]
    \centering
    \includegraphics[width=7.5cm]{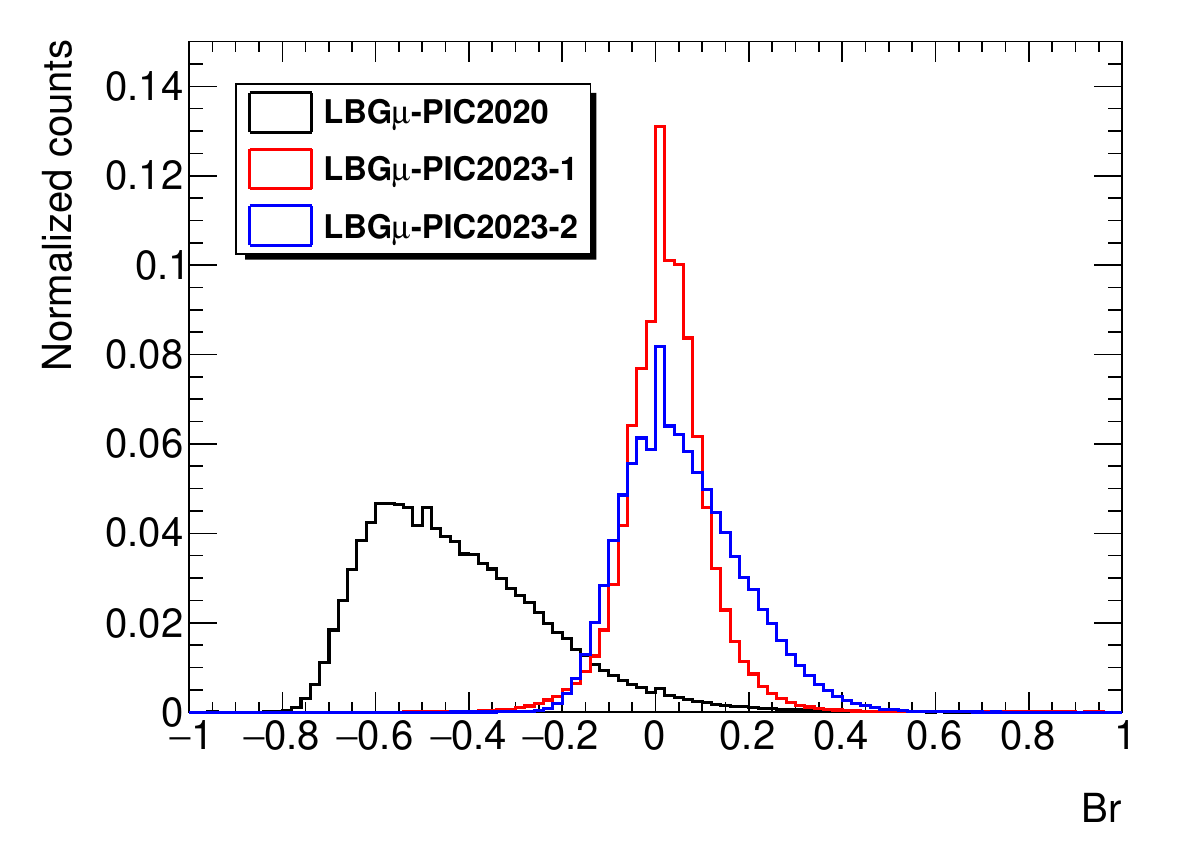}
    \caption{One-dimensional distribution of $Br$. The vertical axis is normalized by the total number of measurements. The mean and the variance values were improved in the production of 2023s.}
    \label{fig:norm1d}
\end{figure}

The two-dimensional and one-dimensional distributions of $Br$ are shown in Fig.~\ref{fig:norm2d} and Fig.~\ref{fig:norm1d}, respectively. LBG\uPIC2020 has a stripe structure in the $Br$ map, indicating that there are regions where the formation of anode electrodes are failed, {\it i.e.} the anode electrodes were well-formed only on the green lines in the map. It implies the non-uniformity of the gas gain, which we will see in Section~\ref{sec:gas_gain_measurement}. The production procedure was improved with this information and the second generation LBG{\uPIC}s were produced in 2023. The $Br$ maps of LBG\uPIC2023-1 and LBG\uPIC2023-2 show a clear improvement of the anode electrodes' formation.

\begin{figure*}[htbp]
    \begin{tabular}{ccc}
        \begin{minipage}{.32\textwidth}
            \centering
            \includegraphics[height=3.8cm]{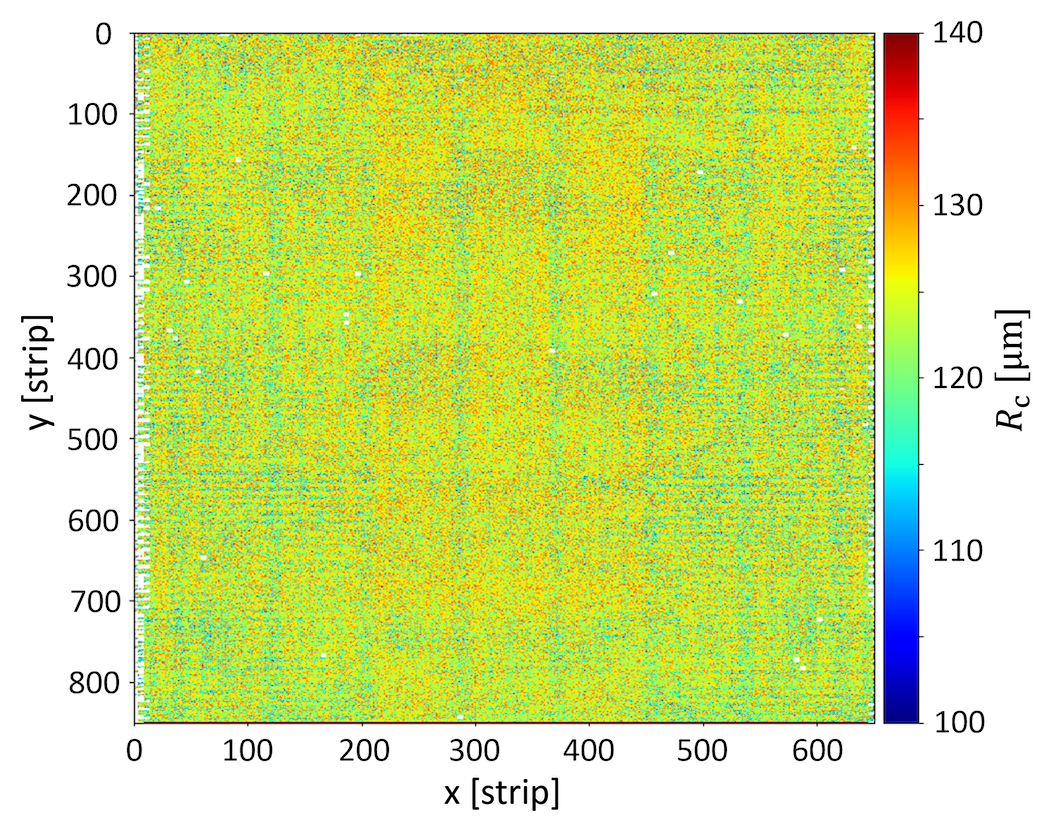}
            \subcaption{LBG\uPIC2020}
            \label{fig:lbg1_radius2d}
        \end{minipage}
        \hspace{.01\textwidth}
        \begin{minipage}{.32\textwidth}
            \centering
            \includegraphics[height=3.8cm]{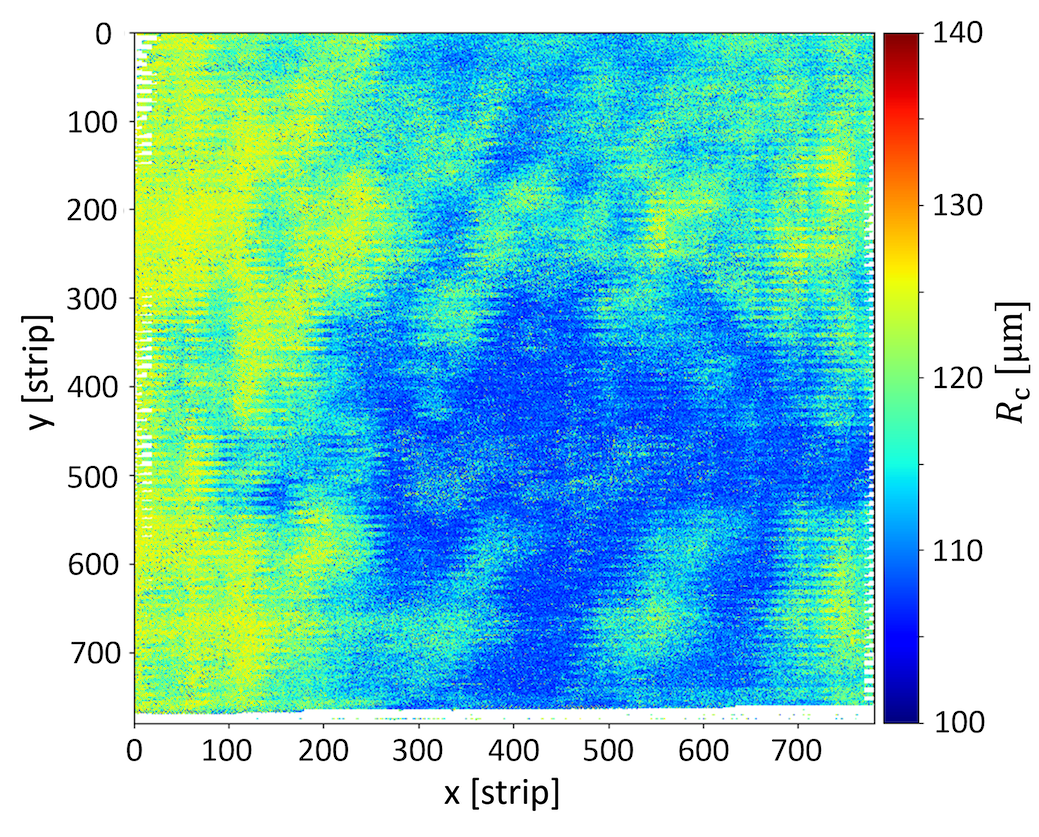}
            \subcaption{LBG\uPIC2023-1}
            \label{fig:lbg3_radius2d}
        \end{minipage}
        \hspace{.01\textwidth}
        \begin{minipage}{.32\textwidth}
            \centering
            \includegraphics[height=3.8cm]{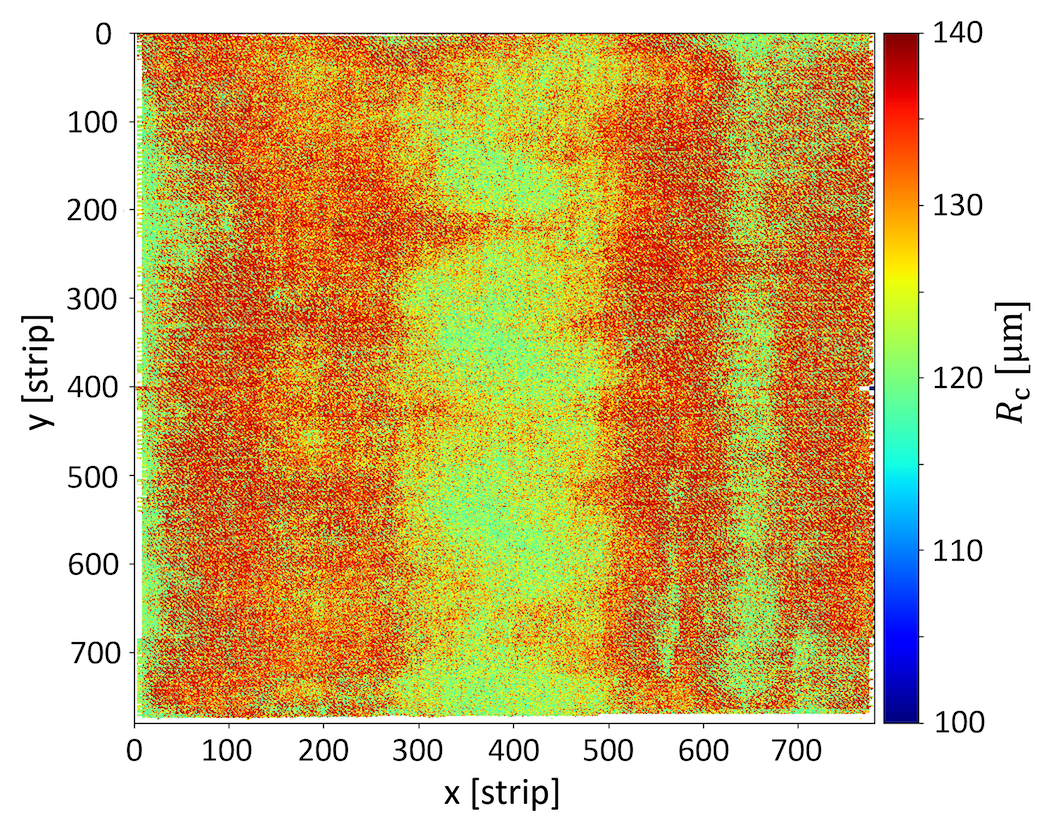}
            \subcaption{LBG\uPIC2023-2}
            \label{fig:lbg4_radius2d}
        \end{minipage}
    \end{tabular}
    \caption{Two-dimensional distributions of $R_{\mathrm{C}}$. 
    }
    \label{fig:radius2d}
\end{figure*}

The two-dimensional and one-dimensional distributions of $R_{\rm C}$  are shown in Fig.~\ref{fig:radius2d}. While LBG\uPIC2020 has a uniform distribution for $R_{\rm C}$, LBG\uPIC2023-1 and LBG{\uPIC}2023-2 have relatively non-uniform distributions. The impact of $R_{\rm C}$ to the gas gain uniformity is discussed in Section~\ref{sec:discussion}.

\subsection{Gas gain measurement}
\label{sec:gas_gain_measurement}
\begin{figure*}[htbp]
    \centering
    \includegraphics[height=13cm]{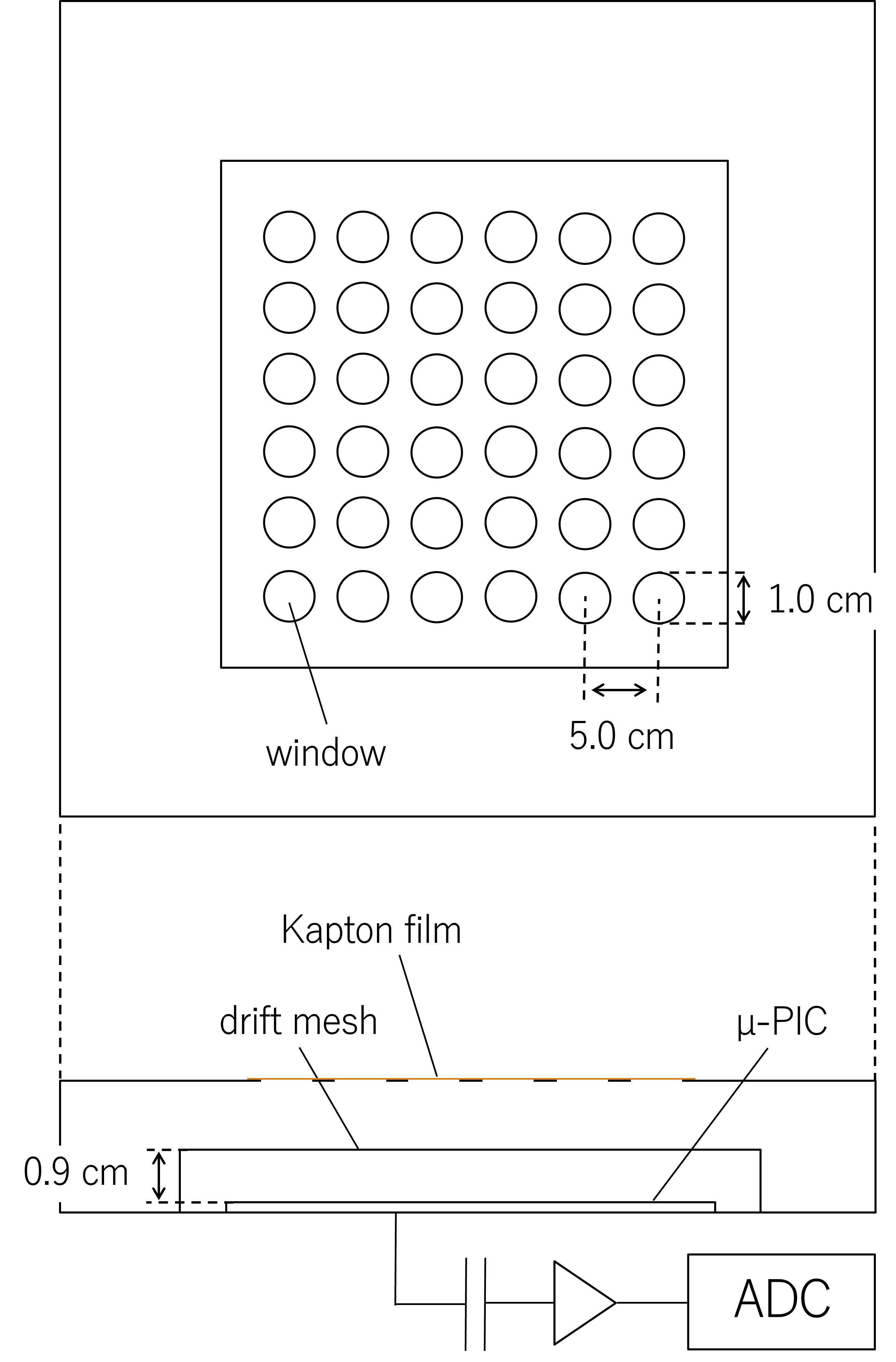}
    \caption{Schematic drawings of the test chamber. It has 6~$\times$~6 holes covered with a sheet of 125~$\mu$m-thick Kapton. }
    \label{fig:detecor_scheme}
\end{figure*}

Gas gains and their uniformities of the LBG{\uPIC}s were measured with a test chamber. Schematic drawings of the test chamber are shown in Fig.~\ref{fig:detecor_scheme}. The test chamber has 6~$\times$~6 holes of $1.0~\mathrm{cm}$ diameter on the top surface with a pitch of $5.0~\mathrm{cm}$. The holes are covered with a sheet of Kapton with a thickness of $125~\mathrm{\mu m}$ to make windows for X-rays. An \elem{55}{Fe} source, which emits $5.9~\mathrm{keV}$ X-rays, was placed on the windows to measure gas gains of the areas of interest. It is experimentally known that the {\uPIC}s show comparable maximum gas gains in $\mathrm{Ar+C_2H_6}$ at $760~\mathrm{Torr}$ as that in $\mathrm{CF_4}$ gas at $76~\mathrm{Torr}$ which is used in the NEWAGE detector. We used one of the standard gas mixtures ($\mathrm{Ar+C_{2}H_{6}}~(9:1)$) at $760~\mathrm{Torr}$ for this measurement. The charge detected by the \uPIC\ was processed with a pre-amplifier (CSA, CR-110, Cremat Inc.) and a shaper (CR-RC Gaussian shaping circuit, two TL074CNs, TI Inc., with a shaping time of $10~\rm{\mu}$sec.). Signals were then digitized by a waveform digitizer (ADALM2000 by Analog Devices, $12~\mathrm{bit~ADC}$, $\pm 2.5~\mathrm{V}$ input range, $10~\mathrm{MHz}$ sampling rate) and stored.

\begin{figure}[htbp]
    \centering 
    \includegraphics[width=7.5cm]{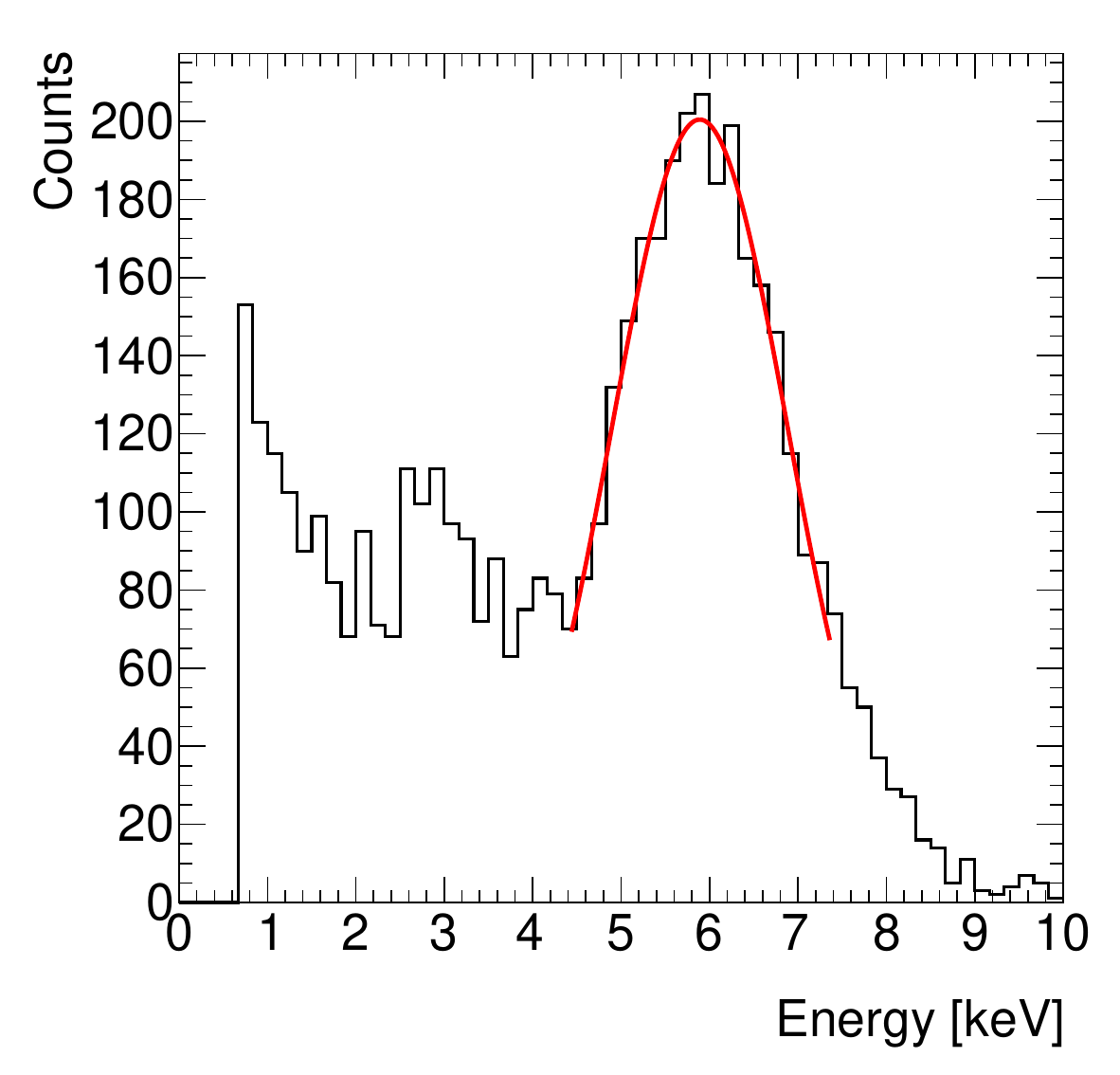}
    \caption{Obtained energy spectrum by the irradiation of X-rays from a \elem{55}{Fe} source to LBG\uPIC2020.}
    \label{fig:spectrum}
\end{figure}

\begin{figure*}[htbp]
    \centering
    \begin{tabular}{cc}
        \begin{minipage}{.49\textwidth}
            \centering
            \includegraphics[height=6cm]{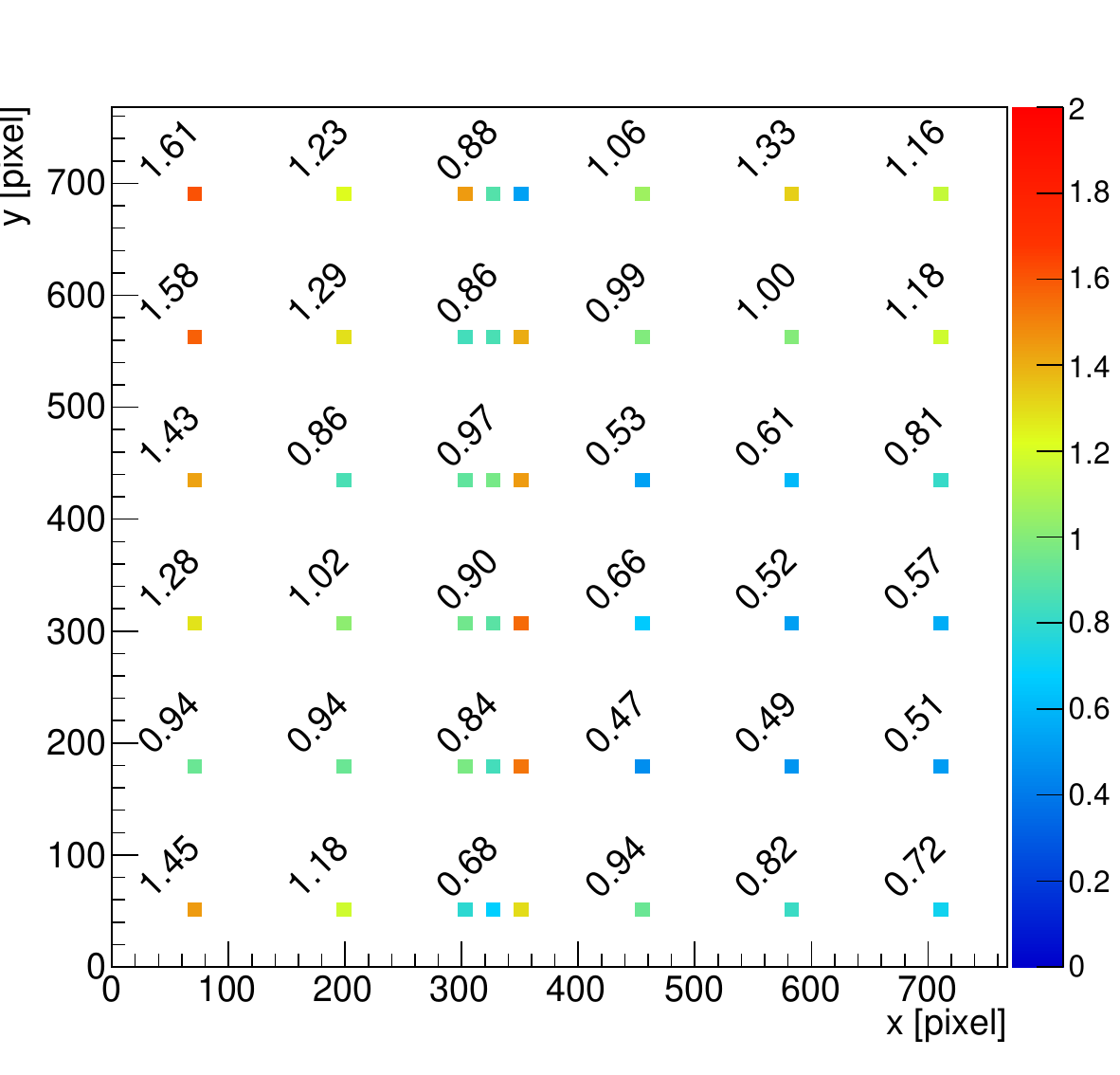}
            \subcaption{LBG\uPIC2020}
            \label{fig:lbg1_gainmap}
        \end{minipage}
        \hspace{.01\textwidth}
        \begin{minipage}{.49\textwidth}
            \centering
            \includegraphics[height=6cm]{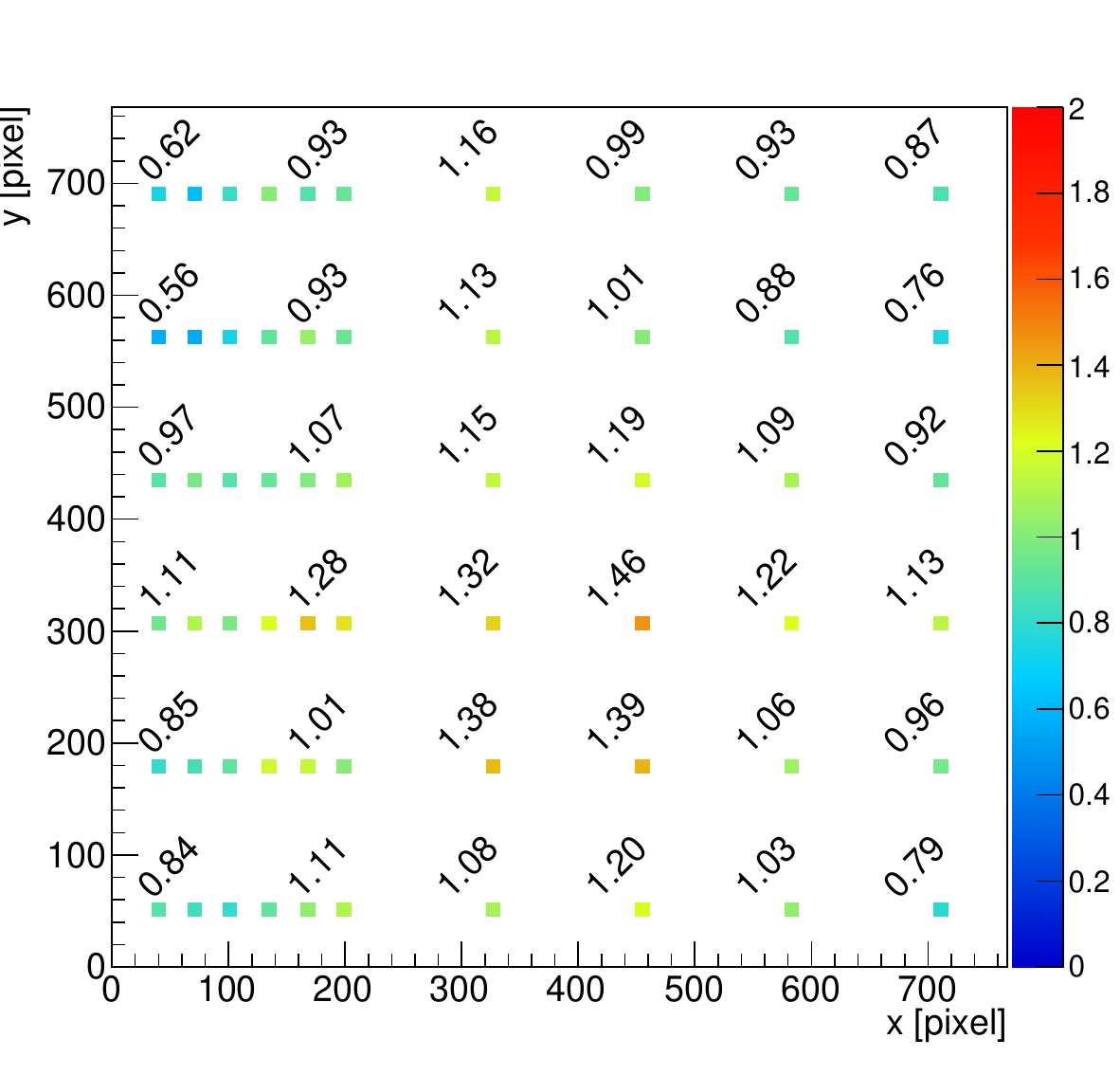}
            \subcaption{LBG\uPIC2023-1}
            \label{fig:lbg3_gainmap}
        \end{minipage}
    \end{tabular}
    \caption{The gain distributions of LBG\uPIC2020 (left) and  LBG\uPIC2023-1 (right). Each gain was obtained by merging neighboring six strips. A local gas gain inhomogeneity was seen in the map of LBG\uPIC 2020. On the other hand, LBG\uPIC2023-1 has no large gain difference within neighboring strips.
    }
    \label{fig:all_gainmap}
\end{figure*}

\begin{figure}[htbp]
    \centering
    \includegraphics[width=7.5cm]{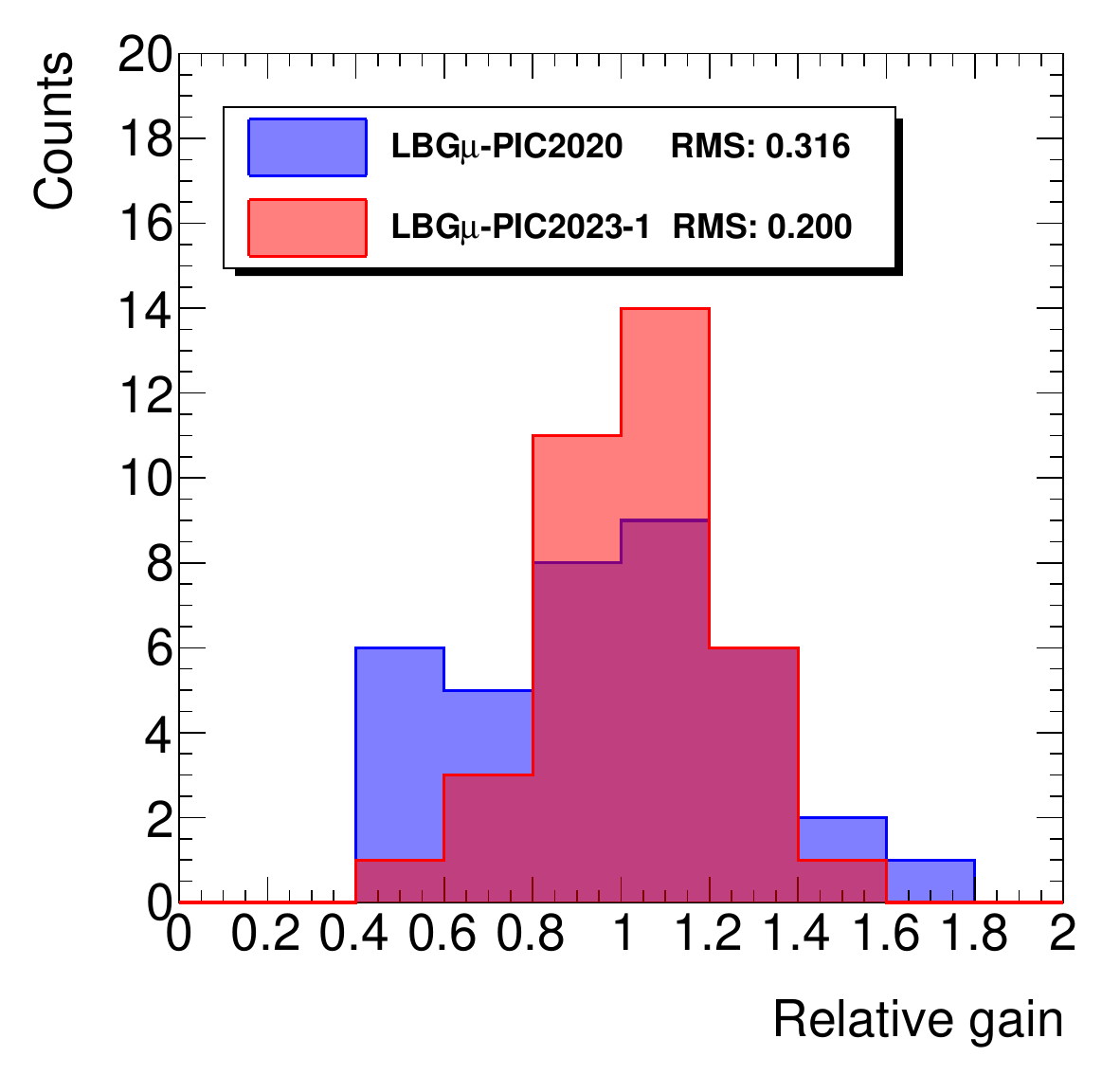}
    \caption{Relative gas gain distributions for LBG\uPIC2020 and LBG\uPIC2023-1 below each Kapton window. The RMS was improved from 31.0\% to 20.0\%.}
    \label{fig:gain_dist1d}
\end{figure}

A typical energy spectrum obtained in this measurement is shown in Fig.~\ref{fig:spectrum}.
Two peaks at 5.9~$\mathrm{keV}$ and 2.9~$\mathrm{keV}$ which correspond to 
the main peak and the escape peak were observed.
The gas gain ($G_{\mathrm{gas}}$) was calculated following the equation
\begin{eqnarray} 
    G_{\mathrm{gas}} & = & \frac{W_{\mathrm{Ar}} \times Q}{5.9\ \mathrm{keV} \times q_{e}},
    \label{eq:gas_gain}
\end{eqnarray}
where $W_{\mathrm{Ar}}$ is the $W$-value of argon (26~eV), $Q$ is the detected charge, and $q_{e}$ is the elementary charge.
Gas gains at various positions for a same voltage were measured. Obtained ``gain maps'' are shown in Fig.~\ref{fig:all_gainmap}. It was found that LBG{\uPIC}2020 had a gain difference of factor two within a few mm distance at some areas ($x\sim$~320 in Fig.~\ref{fig:lbg1_gainmap}). This gain difference in a typical length of the nuclear recoil tracks was problematic for our use. On the other hand, LBG\uPIC2023-1 did not show any gain inhomogeneity in a few mm scale. A large-sale gain inhomogeneity, higher at the center and lower at the edges,  seen in LBG{\uPIC}2023-1 is more easily corrected in the analysis of dark matter searches. Fig.~\ref{fig:gain_dist1d} shows the relative gain distributions for their means below each Kapton window. The overall uniformity has been improved from 30.9\% to 20.0\% in terms of the RMS values. This shows that the requirement for gain uniformity described in Sec.~\ref{sec:requirements} is satisfied.

\begin{figure}[htbp]
    \centering
    \includegraphics[width=8.3cm]{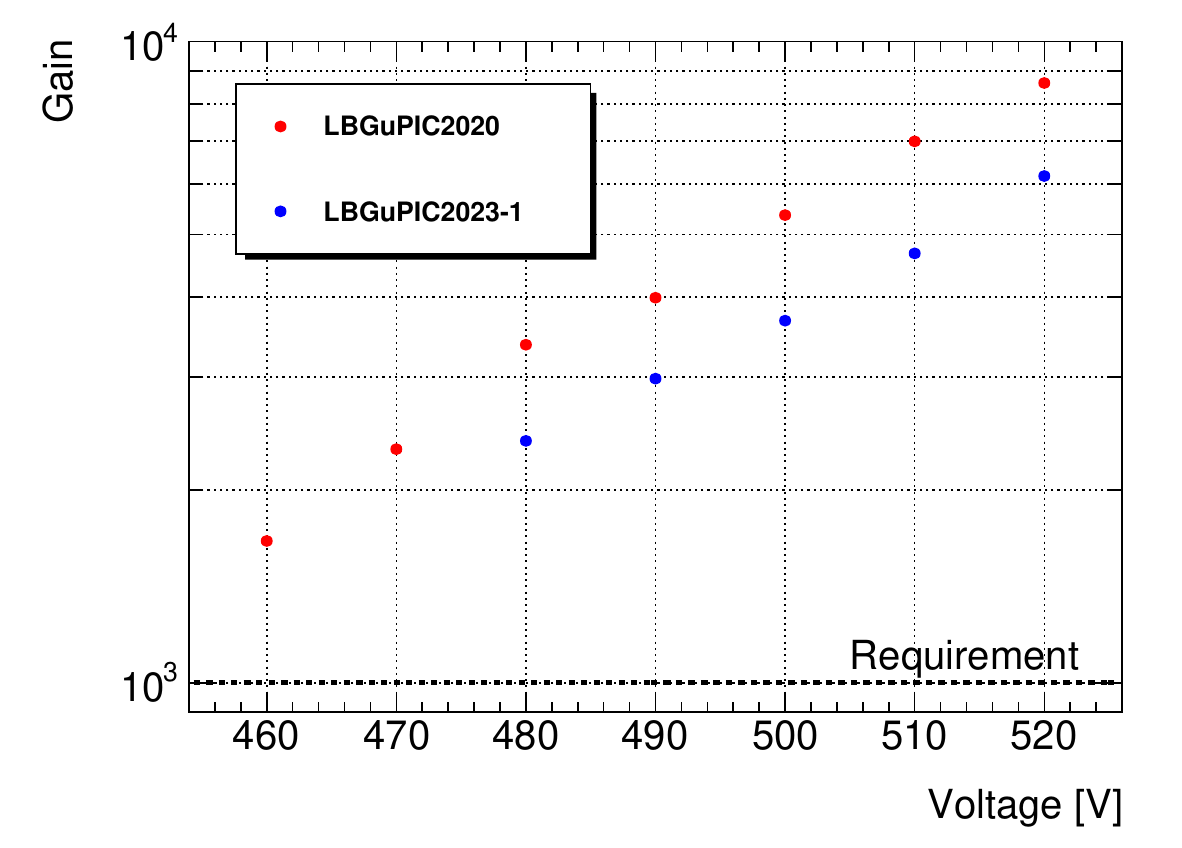}
    \caption{Gain curves for LBG\uPIC2020 and LBG\uPIC2023-1. Gas gains averaged over the detection are shown. Both meet the required values of gain.}
    \label{fig:gain_curve}
\end{figure}

Fig.~\ref{fig:gain_curve} shows 
the gas gains of LBG\uPIC2020 and LBG\uPIC2023-1 averaged over the detection areas as functions of anode voltages. Required gas gain of 1000 was found to be obtained with a voltage higher than 460~V. The gain difference between LBG\uPIC2020 and LBG\uPIC2023-1 of about 20\% will be discussed in Section~\ref{sec:discussion}.

In summary, the requirements of the detector performance were satisfied in terms of the gas gain and its uniformity.

\subsection{Background measurement}
\label{sec:background_measurement}
As discussed in Section~\ref{sec:requirements}, the contamination of \elem{238}{U} and \elem{232}{Th} isotopes in the detector materials will make backgrounds for dark matter searches. In this section, we describe the measurement of the radon emanation and the surface alpha-ray emission from the LBG{\uPIC}s.

\subsubsection{Radon emanation}
\label{sec:radon_emanuation}
Radon emanation rates from the LA\uPIC\ and LBG\uPIC\ were measured at Kobe University with an electrostatic-collection radon detector. This detector was developed based on the method described in Ref.~\citep{Takeuchi1999}. The radon detector has a sample volume of 42~$\times$~42~$\times$~15~$\mathrm{cm^3}$ (27~L), which is large enough to set a \uPIC\ with a size of $\mathrm{41~\times~41~cm^{2}}$. A windowless PIN-photodiode, Hamamatsu S3590-09, was used to measure the energy of alpha rays without any energy loss on the window. The analog signal was digitized and recorded with a waveform digitizer ADALM2000.

The counted radon rates $C$ were fitted with the equation
\begin{equation}
     C = A \times ( 1 - 2 ^ { - t / T } ),
\end{equation}
where $A$ is the radon rate at equilibrium, $t$ is the elapsed time in days, and $T$ is the half-life of \elem{222}{Rn} (3.82~days).

The radon emanation rate was evaluated from the rate of alpha-ray events from \elem{214}{Po} decays. The background rate was also measured and was subtracted from the total event rate. Table \ref{tab:radon_result_sub} shows the \elem{214}{Po} rate for each LBG\uPIC\ after background subtraction.
The \elem{214}{Po} rates were converted to the radon emanation rate as shown in the same table. From this rate, the radon emanation rate of LBG\uPIC2023 was found to be less than 1/60 of that of LA\uPIC.

\begin{table*}[htbp]
    \centering
    \caption{Radon emanation measurement results.  All upper limits are 90\% C.L.}
    \label{tab:radon_result_sub}
    \small
    \begin{tabular}{c|c|c|c}
         Sample & \elem{214}{Po} rate & Radon emanation rate & Radon emanation rate \\ 
         & [count/day] & [$\mathrm{mBq/m^{3}}$] & [mBq/\uPIC] \\
         \hline
         \hline
         LA\uPIC\ & 34.1 $\pm$ 4.9 & 85.2
         $\pm$ 17.4 & 2.3 $\pm$ 0.5 \\  
         \hline
         LBG\uPIC2020 &
         $<$ 2.0 &
         $<$ 5.1 &
         $<$ 0.14 \\ 
         LBG\uPIC2023-1 & $<$ 1.4 & $<$ 1.0 & $<$ 0.03 \\
         LBG\uPIC2023-2 & $<$ 1.3 & $<$ 0.8 & $<$ 0.02  \\
         \hline
    \end{tabular}
\end{table*}

\subsubsection{Surface alpha-ray}
\label{sec:surface_alpha-ray}

We measured the surface alpha-ray emission rate of LBG\uPIC\ with a surface alpha-ray counter (XIA, Ultra-Lo~\cite{UltraLo1,UltraLo2,Nakib:2013ffd,McNally:2014eka}) to evaluate the amount of radon progenies embedding on the final product surface in the manufacturing. The surface alpha emission of the LBG\uPIC\ was found to be $(2.12~\pm~0.28)~\times~10^{-4}~\mathrm{\alpha/cm^{2}/hr}$. This result was consistent with that of LA\uPIC\ ( $(2.35\pm0.48)\times10^{-4}~\mathrm{\alpha/cm^{2}/hr} $) in spite of the material change. Since the measured rate is also consistent with the predictions from the material within error, we can say that the contamination during the manufacturing was negligible.

From the above results, it can be said that the amount of radon emanation rate from the interior was reduced as planed while the surface alpha-ray rate was retained.

\section{Discussion} 
\label{sec:discussion} 

\subsection{Gain estimation by visual inspection}
\label{sec:gain_estimation_by_visual_inspection}

\begin{figure*}[htbp]
    \centering
    \includegraphics[width=15cm]{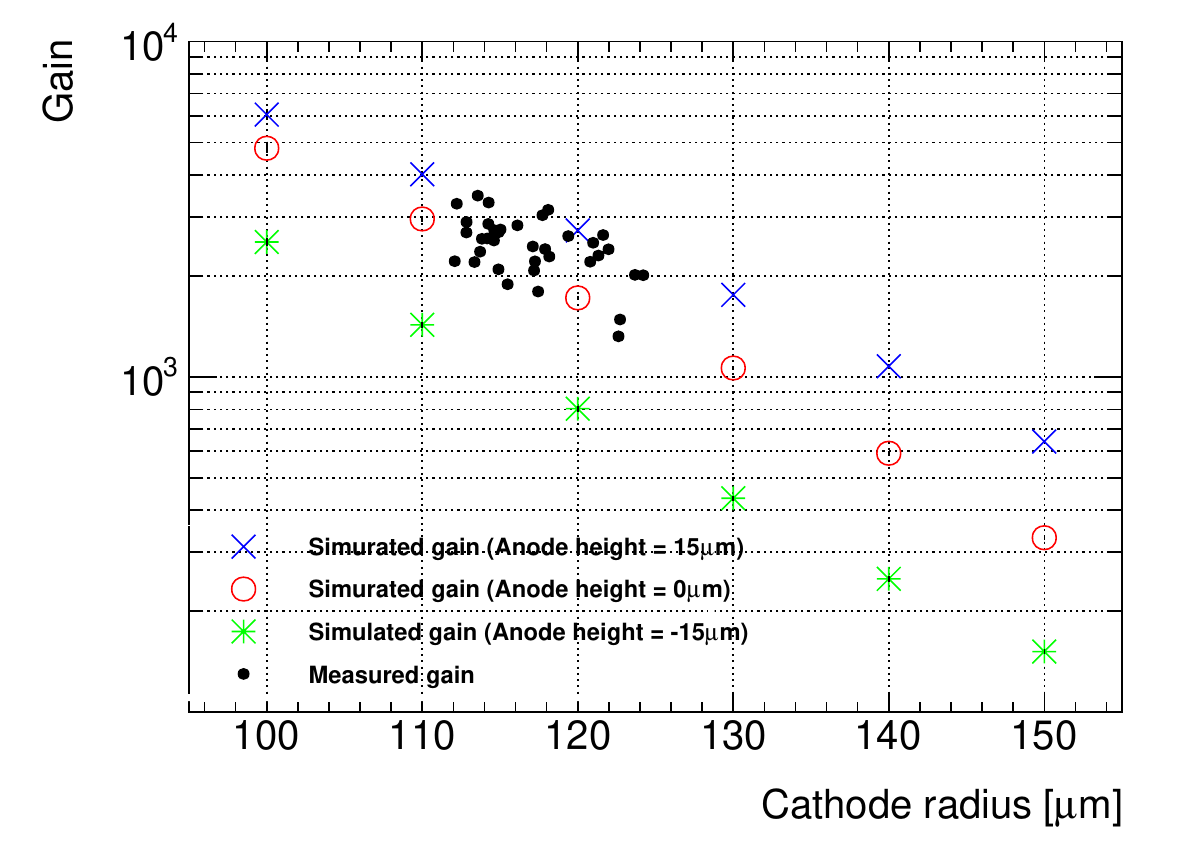}
    \caption{Correlation between cathode radius and gain. The measured gain was obtained from LBG\uPIC2023-1. Red dots represent measured values and blue dots represent gains from Garfield++~\cite{Garfield} simulations. Error bars of simulated gain show the statistic error in simulations.}
    \label{fig:gain_radius}
\end{figure*}

Comparison of Fig.~\ref{fig:lbg3_radius2d} and Fig.~\ref{fig:lbg3_gainmap} indicates a correlation between $R_{\rm C}$ and the gain there. Measured gas gains are shown with black points as a function of $R_{\rm C}$ in Fig.~\ref{fig:gain_radius}. Colored points represent simulated gas gains for given $R_{\rm C}$s and anode heights.  Blue, green, and red markers represents the results with the height of the anode electrodes of -15~$\mathrm{\mu m}$, 0~$\mathrm{\mu m}$, and 15~$\mathrm{\mu m}$, respectively. The trend in the measured data can be explained with the simulation results with the anode height from 0~$\mathrm{\mu m}$ to 15~$\mathrm{\mu m}$. We can see that the gas gain depends both on $R_{\rm C}$ and the anode height. The averaged $R_{\rm C}$ of LBG\uPIC2020 and LBG\uPIC2023-1 were 122.5~$\mathrm{\mu m}$ and 114.8~$\mathrm{\mu m}$, respectively. The larger gas gains of LBG\uPIC2023-1 seen in Fig.~\ref{fig:gain_curve} can be explained by smaller $R_{\rm C}$. From these results, we can say that the uniformity of $R_{\rm C}$ as well as the height of the anode electrodes needs be improved to improve the uniformity of the \uPIC\ gain.

\subsection{Feasibility as a dark matter detector}
\label{sec:feasibility_as_a_dark_matter_detector}
As a dark matter detector component, this LBG\uPIC\ achieved the required lower background level, one-tenth that of  LA\uPIC\ and a sufficient gas gain, over 1000 at the overall detection area. From this background reduction, the expected sensitivity of NEWAGE to SD-WIMP is 1~pb for 100~$\mathrm{GeV/c^2}$ WIMPs with the same exposure as that of Ref.~\cite{Shimada:2023vky}.

\section{Conclusions} 
\label{sec:conclusions}
A micro pixel chamber (\uPIC) with specially selected low background materials (LBG\uPIC) was developed and its performance was studied. The LBG\uPIC\ showed a gas gain more than 1000 with $460-520~\rm{V}$. This gas gain is sufficient for detecting nuclear recoil events in a dark matter search experiment. The radon emanation rate from the LBG\uPIC2023 was found to be less than 1/60 of that from the LA\uPIC. The surface alpha-ray background of the LBG\uPIC\ was the same level as that of the detector components and no contamination during the production from the radons in the air was confirmed. The radon isotope \elem{222}{Rn} emanation from the LBG\uPIC\ was measured. This background reduction of \elem{222}{Rn} from the LBG\uPIC\ could improve the WIMP sensitivity of the NEWAGE.

\section*{Acknowledgments}
\label{sec:acknowledgments}
This work was partially supported by the Japanese Ministry of Education, Culture, Sports, Science and Technology, Grant-in-Aid (19H05806, 26104005, 21H04471, 21K13943, and 22H04574). This work was partially supported by the Inter-University Research Program of the Institute for Cosmic Ray Research (ICRR), the University of Tokyo. We thank the XMASS collaboration, Yoshizumi Inoue, and Yuuki Nakano for their advice on the low-background measurement technologies. 

\bibliographystyle{elsarticle-num}
\bibliography{cas-refs}



\end{document}